\documentclass{elsarticle}
\usepackage{lineno,hyperref}

\usepackage{amssymb}
\usepackage{fancyhdr}
\usepackage[normalsize,bf]{caption}
\usepackage[intlimits]{amsmath}
\usepackage{amsthm}
\newtheorem{rem}{Remark}[section]
\usepackage{stmaryrd}
\usepackage{graphicx}
\usepackage{wrapfig}
\usepackage{bm}  
\usepackage{enumerate}
\usepackage{mathrsfs} 
\usepackage{booktabs}
\usepackage{xcolor}
\definecolor{lightgray}{gray}{0.9}
\usepackage{subfigure}
\usepackage{float}
\usepackage{multirow}
\usepackage{changes}
\definechangesauthor[name=YS,color=brown]{YS}
\definechangesauthor[name=YL,color=blue]{YL}
\definechangesauthor[name=YL,color=purple]{Vahid}
\usepackage[normalem]{ulem}
\usepackage{indentfirst}
\usepackage{todonotes}
\DeclareMathOperator{\meas}{meas}
\modulolinenumbers[5]

\journal{Engineering Fracture Mechanics}

\graphicspath{{Figure_screenshots/}}
\DeclareMathOperator{\tr}{tr}
\DeclareMathOperator{\dev}{dev}
\DeclareMathOperator{\divergence}{div}

\newcommand{\Volume}{{\ooalign{\hfil$V$\hfil\cr\kern0.08em--\hfil\cr}}}
\begin{document}

\begin{frontmatter}

	\title{A MICROMECHANICS-INFORMED PHASE FIELD MODEL FOR BRITTLE FRACTURE ACCOUNTING FOR THE UNILATERAL CONSTRAINT}



	\author[mymainaddress,co1]{Yangyuanchen Liu}
	\author[mymainaddress,co1]{Cheng Cheng}


	\author[mymainaddress]{Vahid Ziaei-Rad}
	
	\author[mymainaddress]{Yongxing Shen\corref{mycorrespondingauthor}}
	\cortext[mycorrespondingauthor]{Corresponding author}
	\ead{yongxing.shen@sjtu.edu.cn}

	\address[mymainaddress]{University of Michigan -- Shanghai Jiao Tong University Joint Institute, Shanghai Jiao Tong University, 800 Dongchuan Road, Shanghai, 200240, China}

	\address[co1]{These authors contributed equally to this work.}
	\begin{abstract}
		We propose a new direction-dependent model for the unilateral constraint involved in the phase field approach to fracture and also in the continuous damage mechanics models. The construction of this phase field model is informed by micromechanical modeling through the homogenization theory, where the representative volume element (RVE) has a planar crack in the center. The proposed model is made closely match the response of the RVE, including the frictionless self-contact condition. This homogenization approach allows to identify a direction-dependent phase field model with the tension-compression split obtained from cracked microstructures. One important feature of the proposed model is that unlike most other models, the material degradation is consistently determined without artificial assumptions or \emph{ad hoc} parameters with no physical interpretation, thus, a more realistic modeling is resulted. With standard tests such as uniaxial loadings, three-point bending, simple shear, and through-crack tests, the proposed model predicts reasonable crack paths. Moreover, with the RVE response as a benchmark, the proposed model gives rise to an accurate stress-strain curve under shear loads, more accurate than most existing models. 
	\end{abstract}


\end{frontmatter}

\linenumbers

\section{Introduction}

The prediction of failure mechanisms due to crack initiation and propagation in solids is of great significance for engineering applications. Griffith's theory \cite{griffith1921} provides a criterion for crack propagation but it is unable to predict crack initiation, merging, and branching. 

Based on the Griffith's theory \cite{griffith1921}, Francfort and Marigo \cite{francfort1998revisiting} proposed a variational theory of fracture and Bourdin \emph{et al.}~\cite{bourdin2000numerical} regularized this theory for numerical computation, the outcome of which is also named the phase field approach to fracture. The phase field modeling of brittle fracture has shown its advantages on simulating complex fracture processes, see \cite{bourdin2000numerical,amor2009regularized,miehe2010phase}.

One of the main challenges in phase field modeling is how to account for the unilateral constraint in material degradation. 
The original model proposed in \cite{bourdin2000numerical} adopts an isotropic response of the cracked solid, i.e., it assumes that both tension and compression loads contribute to cracking. This leads to unphysical crack propagation under a compressive load. 
With the need of direction-dependent material degradation, a number of models have been developed. 
Early models that do distinguish tension vs.~compression were those proposed in \cite{amor2009regularized}, \cite{miehe2010phase}, and \cite{FREDDI20101154}. Amor \emph{et al.} \cite{amor2009regularized} built their model based on the volumetric-deviatoric (V-D) split of the strain tensor, while Miehe \emph{et al.}~\cite{miehe2010phase} based their model on the spectral decomposition of the strain tensor. Around roughly the same time, Freddi and Royer-Carfagni~\cite{FREDDI20101154} proposed a V-D split for masonry-like materials that accounts for the Poisson effect. 
The V-D split \cite{amor2009regularized} and the spectral decomposition \cite{miehe2010phase} are {both} widely used. However, it is well known that in certain tests, unphysical predictions are resulted, e.g. bending test with the V-D split and through-crack shear test with the spectral decomposition.
It is also noted that a nonvariational formulation of the V-D split is proposed in \cite{ambati2015review}. This model preserves the linearity of the symmetric mechanical behavior, and provides more realistic crack propagation at compressive loads, though a modified evolution equation is used for the phase field variable.

Recently, more sophisticated models based on the local crack orientation were proposed. For example, in the two models by Strobl and Seelig \cite{Strobl2015155,strobl2016constitutive},
the crack orientation is assumed to coincide with the local gradient direction of the phase field. Note that both models are variationally inconsistent, thus, they need to introduce \emph{ad hoc} criteria for the evolution of phase field{, as opposed to variationally consistent models following energy minimization}. 
In the model proposed by Steinke and Kaliske \cite{Steinke2018}, a crack coordinate system is {introduced to represent the local crack orientation}, and the split is based on the decomposition of the stress tensor with respect to the {said} crack orientation. Wu \emph{et al.}~\cite{WU2020112629} utilized a positive/ negative projection based on the spectral decomposition to model the unilateral behavior, where the crack orientation is assumed to be the principal direction of the effective stress tensor. This model is expected to share the same disadvantage as those employing the spectral decomposition.


{In summary,} in these models to-date, the crack state is approximately identified by the definition of an artificial condition, e.g., sign of volumetric strain, principal strain/ stress, or crack normal stress. In addition, the decomposition of strains and stresses into crack-driving and persistent portions are somehow artificially introduced into these models, which can lead to predicting unrealistic crack propagation and mechanical behavior.



In 2018,
Cheng \cite{cc2017thesis} and Cheng and Shen \cite{cc2018con} proposed an innovative model for the unilateral constraint involved in the phase field approach {for} brittle fracture for a two-dimensional case. Through homogenization techniques, a link is established between the cracked microstructures and the corresponding macroscopic phase field model. 
For this purpose the authors built the model with inspiration by Dascalu \emph{et al.}~\cite{dascalu2008damage}. 
The main feature of this approach is that as the constitutive model is based on the RVE response,  no artificial condition or material parameter is introduced into the model, unlike for the aforementioned split models. 
In this way, the constitutive model is obtained either with analytical expressions or by numerical fitting.

We note that  independently, {Storm \emph{et al.~}}\cite{kaliske2019} constructed a homogenization framework for their phase field model in a {somewhat} similar way.

This paper is a continuation of \cite{cc2018con}. We present a three-dimensional phase field formulation, which allows {deriving} a variationally consistent direction-dependent material degradation.
More precisely, we model the macroscopic crack as a collection of microscopic cracks, each located in the center of a very small volume element, which is called a representative volume element (RVE), as in the homogenization theory. Then the behavior of each RVE is obtained through a detailed analysis, in particular with the frictionless self-contact condition exactly enforced. The overall behavior of the RVE is then used to construct the phase field model.


The proposed model is subjected to a number of simple tests and compared with a few existing models. In the uniaxial compression test, three-point bending test, and simple shear test, the proposed model behaves similarly as the spectral decomposition model \cite{miehe2010phase}. In the through-crack shear tests, it is well known that the spectral model responds in an unreasonable way, i.e., the material does not break completely across the crack. 
The proposed model avoids such phenomenon, as does the V-D split model \cite{amor2009regularized}.

Another feature is the response of the proposed model in the case of a non-through crack subjected to shear tests. In such tests,
 the proposed model predicts a response very close to the RVE analysis, and it is the only one that provides such uniformly accurate result. One reason that enables such accurate prediction is that the model has not only a tensile term and a compressive term, but also a shear term, of which both the tensile and shear terms will be degraded by the phase field, but to different degrees. While the degradation function for tensile tests is pre-selected, that for shear is obtained by fitting. Moreover, the fitted expression has an explicit dependence on the Poisson ratio.

{One restriction of the model is that the formulation is obtained for isotropic linear elasticity, though generalization to anisotropy or finite elasticity is possible.}

The content of the paper structured as follows. In Section \ref{construction}, a number of existing models are recapitulated. In Section \ref{ourmodel}, the construction of the proposed phase field model is introduced in detail. In Section \ref{numerical}, several numerical examples are tested and results from existing models are compared with the proposed model. It is found that only the proposed model gives an accurate response in all such simple tests, with a benchmark as a more detailed and costly three-dimensional RVE. Finally, a summary of the features of the proposed model is presented in Section \ref{conclusion}.

\section{Existing phase field models}
\label{construction}
In this section, essential ingredients of a few phase field models are recapitulated and compared, with a focus on the tension-compression decomposition.


Most phase field approaches to brittle fracture are based on the variational theory of fracture proposed in \cite{francfort1998revisiting}, which in turn is built on the theory of brittle fracture by \cite{griffith1921}. In the framework of \cite{francfort1998revisiting}, of central importance is the following energy functional:
\begin{equation*}
	\Pi\left ( \bm{u},\Gamma \right ) = \int_{\Omega} \varPsi_{0}\left ( \bm{\varepsilon} ( \bm{u} ) \right )\mathrm{d}\Omega +  \int_{\Gamma}g_{c}\;\mathrm{d}s,
\end{equation*}
where $\Omega \subset \mathbb{R}^n$, $n=2,3$, is the domain occupied by the material at the initial configuration, $\Gamma$ is the crack path, $g_{c}$ is the energy release rate of crack propagation, and $\varPsi_0$ denotes the strain energy density of the (pristine) material, which is a function of the strain tensor $\bm{\varepsilon}(\bm{u})=[\nabla \bm{u} + (\nabla \bm{u})^T]/2$. For an isotropic linear elastic material, the expression of $\varPsi_0$ reads
	\begin{equation*}
	\varPsi_0\left(\bm{\varepsilon}\right)
	= \frac{\lambda}{2} \left ( \tr \bm{\varepsilon}\right )^{2} + \mu \bm{\varepsilon}:\bm{\varepsilon},
	\end{equation*}
where $\lambda$ and $\mu$ are Lam\'{e} constants such that $\mu>0$ and $\lambda+2\mu>0$.

\begin{figure}[htbp] 
	\centering
	\includegraphics[width=\linewidth]{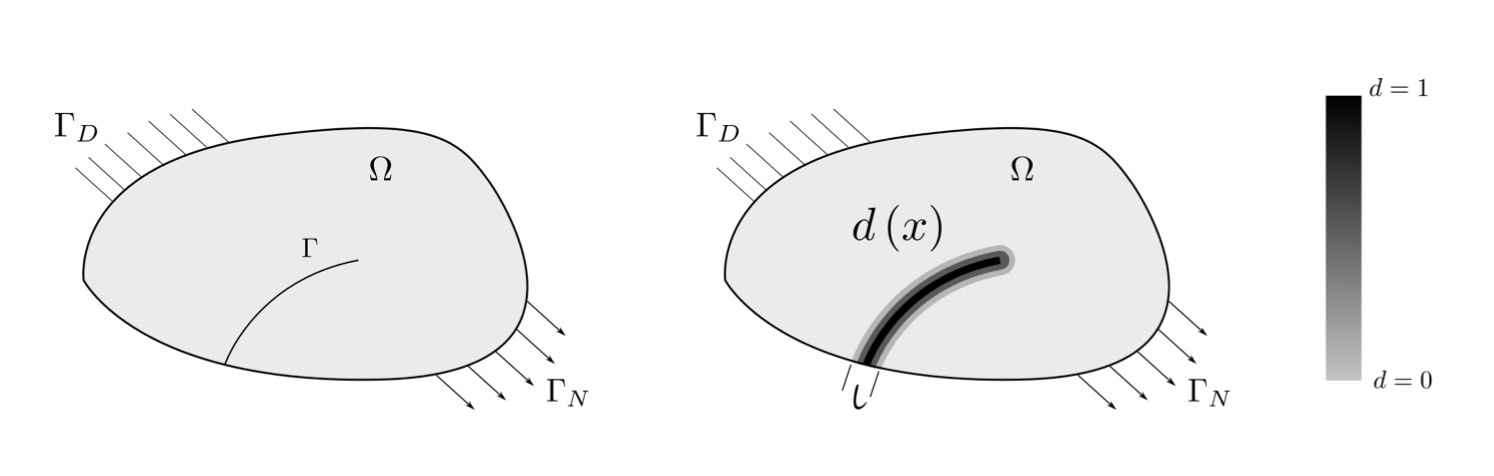}
	\caption{Comparison of a sharp crack and its phase field representation. The left figure illustrates the sharp crack model; 
		the right figure shows the phase field representation of the crack through a smooth transition area, in which the phase field variable is between 0 and 1. The phase field is one of the arguments in the regularized variational theory of fracture, also named the phase field approach to fracture.
	}
	\label{phasefield_illus}
\end{figure}

 
In order to construct a numerical simulation method for the variational theory of fracture, 
\cite{bourdin2000numerical} put forward a regularized formulation, in which the functional $\Pi$ is regularized as $\Pi_l$: 
\begin{equation*}
\Pi_{l}\left [ \bm{u},d \right ]:= \int_{\Omega}\varPsi\left [\bm{\varepsilon}\left ( \bm{u} \right ),d \right ]\mathrm{d}\Omega + \frac{g_{c}}{2}\int_{\Omega}\left ( \frac{d^2}{l} + l\left | \nabla d \right |^2 \right )\mathrm{d}\Omega,
\end{equation*}
which is a functional of the displacement field $\bm{u}$ and the phase field $d$, see Figure \ref{phasefield_illus}. Here $l$ is a regularization length scale parameter. Note that for $d$ we adopt a convention \emph{opposite} to that of \cite{bourdin2000numerical}, i.e., $d=1$ in this manuscript represents the crack and $d=0$ pristine material.

For generality, we let $\bm{t}_{N}$ denote the traction field applied on $\Gamma_N \subset \partial\Omega$ and $\bm{b}$ the body force field, then the expression of $\Pi_l$ is modified to be
\begin{equation}\label{variational}
\Pi_{l}\left [ \bm{u},d \right ]= \int_{\Omega}\varPsi\left [\bm{\varepsilon}\left ( \bm{u} \right ),d \right ]\mathrm{d}\Omega - \int_{\varGamma_{N}}\bm{t}_{N}\cdot\bm{u}\;\mathrm{d}\varGamma - \int _{\Omega}\bm{b}\cdot\bm{u}\;\mathrm{d}\Omega + \frac{g_{c}}{2}\int_{\Omega}\left ( \frac{d^2}{l} + l\left | \nabla d \right |^2 \right )\mathrm{d}\Omega.
\end{equation}%


We particularly focus on the strain energy density term $\varPsi\left (\bm{\varepsilon}, d \right )$. Once $\varPsi$ is decided, the Cauchy stress tensor is given by
\begin{equation*}
\bm{\sigma}(\bm{\varepsilon},d) = \frac{\partial \varPsi}{\partial \bm{\varepsilon}},
\end{equation*}
and the driving force for the evolution of $d$ is given by $-\partial\varPsi/\partial d$.

The requirements on $\varPsi$ are:
\begin{enumerate}
	\item $\varPsi(\bm{\varepsilon}, d=0) = \varPsi_0(\bm{\varepsilon})$.
	\item $\varPsi(\bm{\varepsilon}, d)\ge 0$ for all $\bm{\varepsilon}$ and $d$.
	\item $\partial\varPsi(\bm{\varepsilon}, d)/\partial d \le 0$, for all $\bm{\varepsilon}$ and $d$.
\end{enumerate}

\subsection{Models not taking into account the local crack orientation}
\paragraph{General expression of $\varPsi(\bm{\varepsilon}, d)$}
In most models to follow, $\varPsi(\bm{\varepsilon}, d)$ takes the following form:
\[\varPsi(\bm{\varepsilon}, d) = g(d) \varPsi_+(\bm{\varepsilon}) + \varPsi_-(\bm{\varepsilon}),
\]
where $\varPsi_+$ and $\varPsi_-$ are the crack-driving and persistent parts of the strain energy density, respectively, and $g(d)$ is the degradation function, often taken as $g(d)=(1-d)^2 + k$, with $k$ a small positive number merely to prevent the tangent stiffness matrix of the system from being singular.

The Cauchy stress is thus given by
\begin{equation*}
\bm{\sigma}(\bm{\varepsilon},d) = g(d)\bm{\sigma}_+ + \bm{\sigma}_-, \quad \bm{\sigma}_\pm := \frac{\partial \varPsi_\pm}{\partial \bm{\varepsilon}}.
\end{equation*}

The evolution equation for the phase field variable follows from the minimization principle of $\Pi_l$, i.e.,
\begin{equation}\label{Eq:evolution_d}
\frac{\partial\varPsi}{\partial d}+g_c\left(\frac{d}{\ell}-\ell\nabla d\right)=0.
\end{equation}
The split models constructed on the basis of this evolution equation are variationally consistent.

We next proceed to examine three models.

\paragraph{The isotropic model proposed by \cite{bourdin2000numerical}}
In the formulation of \cite{bourdin2000numerical}, the degraded strain energy density $\varPsi$ is expressed as the entire strain energy density of the unbroken material $\varPsi_0$ multiplied by the degradation function $g(d)$. More precisely,
\begin{align*}
	\varPsi(\bm{\varepsilon},d) &= g(d)\varPsi_0( \bm{\varepsilon}),\\
	\bm{\sigma}(\bm{\varepsilon},d) &= g(d)[\lambda (\tr\bm{\varepsilon})\bm{1} + 2\mu\bm{\varepsilon}],\\
	\frac{\partial\varPsi}{\partial d} &= g'(d)\varPsi_0( \bm{\varepsilon}).
\end{align*}

As well known, since the degradation function is applied to the entire $\varPsi_0$, this model does not distinguish the fracture behavior between tension and compression, resulting in unphysical results when the crack is subjected to compressive or shear loads.

%

\paragraph{The V-D (volumetric-deviatoric) decomposition model proposed by \cite{amor2009regularized}} This model splits the strain energy density into volumetric and deviatoric parts: $\varPsi_0(\bm{\varepsilon}) = \varPsi_{+}( \bm{\varepsilon}  ) + \varPsi_{-}( \bm{\varepsilon}) $, where
\begin{equation*}	
\begin{aligned}
\varPsi_{+}( \bm{\varepsilon}) &= \frac{K}{2}\langle \tr\bm{\varepsilon} \rangle_{+}^2 + \mu\left \| \dev\bm{\varepsilon} \right \|^2, \\
\varPsi_{-}( \bm{\varepsilon}) &= \frac{K}{2} \langle \tr \bm{\varepsilon} \rangle_{-}^2,\\
\bm{\sigma}( \bm{\varepsilon},d ) &= g(d)\left( K \langle \tr\bm{\varepsilon} \rangle_{+} \bm{1} + 2\mu \dev\bm{\varepsilon} \right) + K\langle \tr\bm{\varepsilon}\rangle_{-}\bm{1},\\
\frac{\partial\varPsi}{\partial d} &= g'(d)\left(\frac{K}{2}\langle \tr\bm{\varepsilon} \rangle_{+}^2 + \mu\left \| \dev\bm{\varepsilon} \right \|^2\right),
\end{aligned}
\end{equation*}
where $K = \lambda + 2\mu/3$ is the bulk modulus, $\dev\bm{\varepsilon}:=\bm{\varepsilon} - (1/3)(\tr \bm{\varepsilon})\bm{1}$, and $\langle a \rangle_{\pm}:=(a \pm | a  | )/2$. This split is based on the assumption that the strain energy resulting in the decrease of the local volume will not be degraded.    

\paragraph{The spectral decomposition model proposed by \cite{miehe2010_1273}}  In this model, the split is based on the spectral decomposition of the strain tensor. Let $\left\{ \varepsilon_{a} \right\}^3_{a=1}$ be the principal strains and $\left\{\bm{\mathrm{E}}_{a}\right\}^{3}_{a=1}$ be the corresponding orthonormal principal directions.  The expressions are as follows:
\begin{equation*}
\begin{aligned}
\varPsi_{\pm}( \bm{\varepsilon} ) &= \frac{\lambda}{2}\langle \tr \bm{\varepsilon} \rangle_{\pm}^2 + \mu \sum_{a=1}^{3}\langle \varepsilon_{a}\rangle_{\pm}^2,\\
\bm{\sigma}_{\pm}\left(\bm{\varepsilon}\right) &= \lambda\langle \tr\bm{\varepsilon} \rangle_{\pm}\bm{1} + 2\mu \sum_{a=1}^{3} \langle \varepsilon_{a} \rangle_{\pm} \bm{\mathrm{E}}_{a} \otimes \bm{\mathrm{E}}_{a},
\\
\frac{\partial\varPsi}{\partial d} &= g'(d) \left(\frac{\lambda}{2}\langle \tr \bm{\varepsilon} \rangle_+^2 + \mu \sum_{a=1}^{3}\langle \varepsilon_{a}\rangle_+^2\right).
\end{aligned}
\end{equation*}

\paragraph{Related models in damage mechanics}
The last two splits can trace their roots in the damage mechanics community. For example, the split adopted by in \cite{amor2009regularized} is similar to that by \cite{lemaitre1992}, while the decomposition proposed in \cite{miehe2010_1273} is similar to the formulations of \cite{ortiz1985,simo1987,yazdani1988,lubarda1994}.

\paragraph{The anisotropic model proposed by \cite{WU2020112629}}
In this model, Wu et al. utilized a positive/negative projection based on the spectral decomposition. The spectral decomposition of the effective stress is
\begin{equation*}
	\begin{aligned}
		\overline{\bm{\sigma}} &= \lambda (\tr \bm{\varepsilon}) \bm{1} + 2\mu \bm{\varepsilon},\\
		\overline{\bm{\sigma}} &=\sum_{i=1}^{3} \overline{\sigma}_i\bm{p}_i\otimes\bm{p}_i=\overline{\bm{\sigma}}^+ + \overline{\bm{\sigma}}^- ,\\
		\overline{\bm{\sigma}}^{\pm} &= \sum_{i=1}^{3} \overline{\sigma}_i^{\pm}\bm{p}_i\otimes\bm{p}_i,
	\end{aligned}
\end{equation*}
where $\overline{\sigma}_i$ and $\bm{p}_i$ represent the $i$th, $i=1,2,3$, eigenvalue and eigenvector of the effective stress $\overline{\bm{\sigma}}$, respectively. The principal values $\overline{\sigma}_i^{+}$ are determined as 
\begin{equation*}
	\begin{aligned}
		\overline{\sigma}_1^{+} &= \langle \overline{\sigma}_1 \rangle,\\
		\overline{\sigma}_2^{+} &= \langle \max(\overline{\sigma}_2, \tilde{\nu}\overline{\sigma}_1) \rangle,\\
		\overline{\sigma}_3^{+} &= \langle \max[\max(\overline{\sigma}_3, \nu(\overline{\sigma}_1+\overline{\sigma}_2)), \tilde{\nu}\overline{\sigma}_1] \rangle,
	\end{aligned}
\end{equation*}
where $\tilde{\nu} := \nu/(1-\nu)$.

\subsection{Models with the local crack orientation taken into consideration}


\paragraph{The SS models proposed by \cite{Strobl2015155,strobl2016constitutive}} 
\cite{Strobl2015155,strobl2016constitutive} proposed two models, which we will refer to as SS1 and SS2 models.
In these models, the expression of $\bm{\sigma}$ in terms of $\bm{\varepsilon}$ depends on the local crack orientation, which could be either the ``active'' case or the ``passive'' case.
The local crack orientation is given by $\bm{n} = \nabla d/|\nabla d |$. Also the authors define $\bm{N} = \bm{n}\otimes \bm{n}$.
Both models take the following form:
\[
\bm{\sigma} = \begin{cases}
\bm{\sigma}_\mathrm{act}, & \text{if } \bm{\varepsilon}:\bm{N}>0, \\
\bm{\sigma}_\mathrm{pas}, & \text{if } \bm{\varepsilon}:\bm{N}<0.
\end{cases}
\]

The SS1 model reduces the stiffness parallel to the crack in both active and passive cases, which reads
\begin{align*}
	\bm{\sigma}_\mathrm{act} &=
	g(d) \left( \lambda \tr(\bm{\varepsilon})\bm{1} + 2\mu\bm{\varepsilon} \right),\\
	\bm{\sigma}_\mathrm{pas} &= 
	g(d) \lambda \tr(\bm{\varepsilon})\bm{1} + 2g(d)\mu\bm{\varepsilon} + (1-g(d))(\lambda + 2\mu)\left( \bm{\varepsilon} : \bm{N} \right) \bm{N}.
\end{align*}

The SS2 model only degrades the stiffness normal to the crack. Its expression is
\begin{align*}
	\bm{\sigma}_\mathrm{act} &= 
	\left( \lambda + (g(d) - 1) {\lambda^2 \over {\lambda + 2\mu}} \right) (\tr\bm{\varepsilon})\bm{1} + 2\mu\bm{\varepsilon} \\
	&\quad+ (g(d) - 1) \left( \lambda + \frac{\lambda^2}{\lambda + 2\mu} \right) \left( (\tr\bm{\varepsilon})\bm{N} + (\bm{\varepsilon}:\bm{N}) \bm{1} \right) \\
	&\quad+ 4(1 - g(d))\left(\lambda + 2\mu - \frac{\lambda^2}{\lambda + 2\mu}\right) (\bm{\varepsilon}:\bm{N})\bm{N} + \mu(g(d) - 1)\left(\bm{N}\cdot\bm{\varepsilon} + \bm{\varepsilon} \cdot \bm{N }\right),\\
	\bm{\sigma}_\mathrm{pas} &= 
	\lambda \tr(\bm{\varepsilon})\bm{1} + 2\mu\bm{\varepsilon} + 4\mu(1-g(d))(\bm{\varepsilon}:\bm{N})\bm{N} + \mu(g(d) - 1)\left( \bm{N}\cdot \bm{\varepsilon} + \bm{\varepsilon} \cdot \bm{N}\right).
\end{align*}

A comment for the SS2 model goes as follows. For a uniaxial tension load $\varepsilon_{22}>0$, $\varepsilon_{11}=0$, and $\bm{n}=\bm{e}_2$, where $\bm{e}_i$ is the $i$th Cartesian coordinate axis, when $d=1$,
\begin{align*}
(\bm{\sigma}_\mathrm{act})_{22} = 
\biggl\{
 \lambda - {\frac{\lambda^2}{\lambda + 2\mu}}   
 - 2 \left( \lambda + {\frac{\lambda^2}{\lambda + 2\mu}} \right)  
+ 4\left(\lambda + 2\mu - {\frac{\lambda^2}{\lambda + 2\mu}}\right)   
\biggr\} \varepsilon_{22},
\end{align*}
which is not positive for a certain range of $\lambda/\mu$, giving rise to an unstable material.


\paragraph{The SK model proposed by \cite{Steinke2018}} In this model, the split is based on the decomposition of the stress tensor with respect to the crack orientation. For each point, a crack coordinate system is defined with mutually orthonormal vectors $\bm{n}$, $\bm{s}$, and $\bm{t}$. The crack orientation $\bm{n}$ is obtained from the maximum principal stress direction in this model.
The expressions of the directional split are as follows: 
\begin{align*}
	\bm{\sigma} &= g(d)\bm{\sigma}^+ + \bm{\sigma}^-, \\
	\bm{\sigma}^+ &= H(\sigma_{nn})\left[\bm{n}\otimes\bm{n}+ \frac{\lambda}{\lambda+2\mu}(\bm{s}\otimes\bm{s}+\bm{t}\otimes\bm{t})\right] (\lambda \bm{1}+2\mu\bm{n}\otimes\bm{n}):\bm{\varepsilon} \\
	&+ \mu[(\bm{n}\otimes\bm{s}+\bm{s}\otimes\bm{n})\otimes(\bm{n}\otimes\bm{s}+\bm{s}\otimes\bm{n})+(\bm{t}\otimes\bm{n}+\bm{n}\otimes\bm{t})\otimes(\bm{t}\otimes\bm{n}+\bm{n}\otimes\bm{t})]:\bm{\varepsilon},\\
	\bm{\sigma}^- &= \left[ H(-\sigma_{nn}) \bm{n}\otimes\bm{n} - H(\sigma_{nn}) \frac{\lambda}{\lambda+2\mu} (\bm{s}\otimes\bm{s} + \bm{t}\otimes\bm{t})\right] \otimes (\lambda \bm{1}+2\mu\bm{n}\otimes\bm{n}):\bm{\varepsilon}\\
	&+ (\bm{s}\otimes\bm{s})\otimes(\lambda \bm{1}+2\mu\bm{s}\otimes\bm{s}) : \bm{\varepsilon} + (\bm{t}\otimes\bm{t})\otimes(\lambda \bm{1}+2\mu\bm{t}\otimes\bm{t}) : \bm{\varepsilon}\\
	&+ \mu[(\bm{s}\otimes\bm{t}+\bm{t}\otimes\bm{s})\otimes(\bm{t}\otimes\bm{s}+\bm{s}\otimes\bm{t})]:\bm{\varepsilon},
\end{align*}
where the degradation function $g(d;b)$ is given by  
 \begin{equation*}
 g(d;b)=\frac{\exp(b d) - (b (d-1)+1) \exp(b)}{(b-1)\exp(b)+1},
 \end{equation*}
 where $b$ is an extra parameter to be specified.
 

\paragraph{Remarks}
All the aforementioned models except the SS models
are variationally consistent, while the SS models are variationally inconsistent.  
With the relaxation of variational consistency, the efficiency of the numerical calculation may be improved. Nevertheless, non-variational models give rise to additional complications, in that they need to introduce extra damage or failure criteria for the evolution of $d$. In contrast, variationally consistent models are closely related to Griffith's theory and would better fit in a thermodynamic framework. 
Hence, the proposed model will also be made variationally consistent.



\section{The proposed micromechanics-informed phase field model}
\label{ourmodel}
In this section, the steps for constructing the proposed phase field model are detailed.

In order to construct our model, we borrow the basic concepts from Dascalu \emph{et al.~}\cite{dascalu2008damage}, in which the authors applied asymptotic homogenization techniques to {model} the overall behavior of a damaged elastic body. The main ingredient is a micromechanical energy analysis, performed on a finite-sized cell, which leads, through homogenization, to a macroscopic evolution equation for {the damage variable}. Here, we follow the same idea to obtain a three-dimensional phase field model that incorporates {direction-dependent} material degradation.

\begin{figure}[htbp!]
	\centering
	\includegraphics[width=\linewidth]{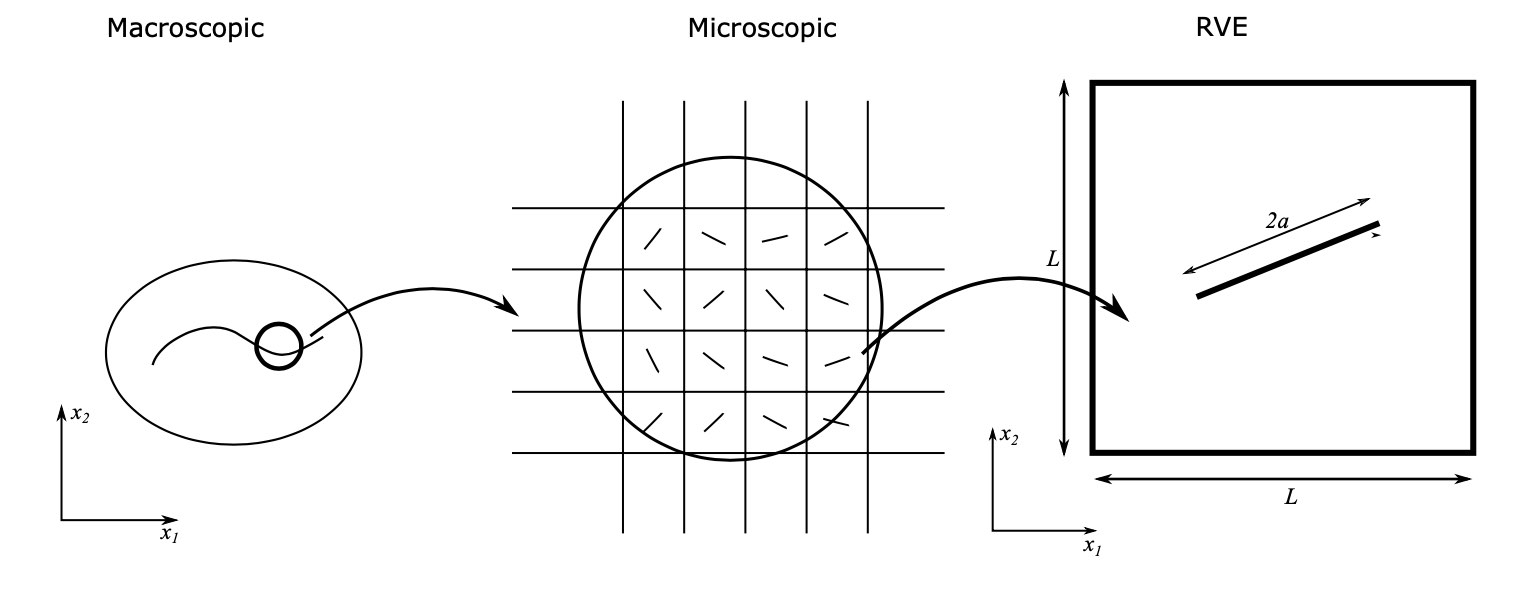}
	\caption{Modeling a macroscopic crack as a collection of fictitious small straight cracks (planar crack in 3D), each in one RVE.}
	\label{RVE}
\end{figure}

Essentially we model a possibly curved crack as a collection of fictitious small flat cracks with a particular spatial distribution. More precisely, we assume that a portion of the solid can be divided into many cube-shaped subdomains (or square-shaped in 2D), at the center of each of which there exists a said small planar crack (or straight crack in 2D), similar to the configuration proposed in \cite{dascalu2008damage}. The behavior of the macroscopic crack is then described by the collection of such small cracks, see Figure \ref{RVE}. Here we borrow a terminology of the homogenization theory and call each cube-shaped subdomain a representative volume element (RVE), and will apply standard homogenization techniques to extract its response subjected to mechanical loading.

In a nutshell, we will establish a one-to-one relation between the crack dimension in the RVE and the local phase field $d$. Then for any macroscopic input strain tensor $\overline{\bm{\varepsilon}}\in\mathbb{R}^{n\times n}$ ($n=2,3$) with standard homogenization technique, we can calculate an average stress $\overline{\bm{\sigma}}\in\mathbb{R}^{n\times n}$, and then obtain $\mathbb{C}\in\mathbb{R}^{n\times n\times n\times n}$, the effective secant modulus, such that
\begin{equation}\label{psi-macro}
\overline{\bm{\sigma}}(\overline{\bm{\varepsilon}}, d) = \mathbb{C}(\overline{\bm{\varepsilon}},d) : \overline{\bm{\varepsilon}}.
\end{equation}


The expression of $\mathbb{C}(\overline{\bm{\varepsilon}}, d)$ can be obtained by comparison with analytical expressions or by numerical fitting. This way $\varPsi(\overline{\bm{\varepsilon}}, d)$ can also be obtained. 

\subsection{General idea of the homogenization approach}
\begin{figure}[h]
	\centering
	\includegraphics[width=\linewidth]{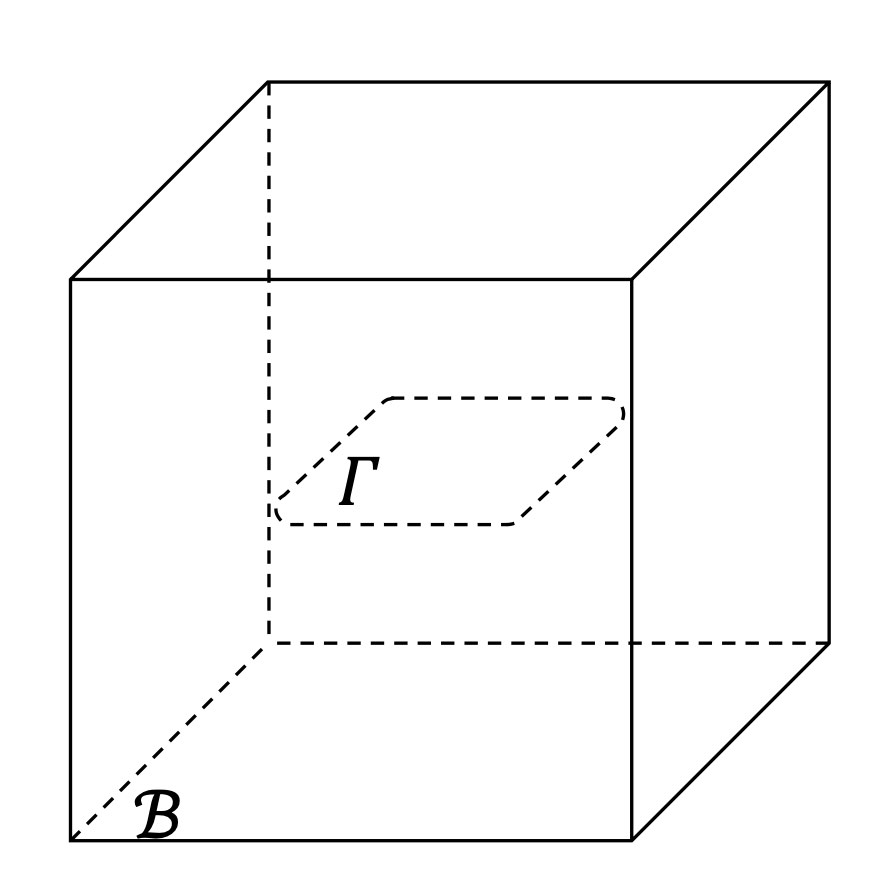}
	\caption{The boundary value problem of RVE}
	\label{rvebc}
\end{figure}

Consider the RVE shown in Figure \ref{rvebc}.
Let the material be isotropic elastic with Lam\'{e} parameters $\lambda$ and $\mu$ such that $\mu>0$ and $\lambda+2\mu>0$. 
We denote by $\mathcal{B}$ the domain occupied by the RVE, a bounded open subdomain of $\mathbb{R}^n$ with a smooth boundary.
Without loss of generality, let $\varGamma\subset\mathcal{B}$ be a micro-crack normal to the $x_2$-axis. In the sequel we will test three types of micro-cracks: (i) a straight crack at the center of a 2D RVE; (ii) a penny-shaped crack at the center of a 3D RVE; (iii) a square-shaped crack with rounded fillets ($r=0.05L$) at the center of the 3D RVE, as shown in Figure \ref{rvebc}. Details of the numerical experiments on the RVE are provided in \ref{sec:Abaqus}.


The computational domain is then $\mathcal{B}_s = \mathcal{B}\setminus\varGamma$.
In $\mathcal{B}_s$, the equilibrium equation and boundary conditions are
\begin{subequations}\label{RVE-BVP}
\begin{align}
\divergence \bm{\sigma}  = \bm{0}, \quad&\text{in } \mathcal{B}_s,&\label{RVEeqm}
\\
\label{constitutive}
\bm{\sigma} = \mathbb{C}: \nabla \bm{u}, \quad&\text{in } \mathcal{B}_s,&
\\
\label{boundary}
\bm{u} = \overline{\bm{\varepsilon}}\cdot \bm{x}, \quad&\text{on } \partial\mathcal{B},& 
\\\label{contact}
\left.
\begin{aligned}
\text{either }
\bm{\sigma}\cdot \bm{n} = \bm{0}, \llbracket\bm{u}\cdot\bm{n}\rrbracket > 0,\\
\text{or }		\llbracket\bm{\sigma}\cdot \bm{n}\rrbracket = \bm{0}, \bm{n}\cdot \bm{\sigma}\cdot\bm{n} < 0, \bm{t}\cdot \bm{\sigma}\cdot\bm{n} = \bm{0}, \llbracket
\bm{u}\cdot\bm{n}\rrbracket = 0,\end{aligned}
\right\}
& \quad\text{on } \varGamma,&	\end{align}
\end{subequations}
where \eqref{contact} is the frictionless contact condition for $\varGamma$, 
$\bm{n}$ and $\bm{t}$ are the unit normal and tangent vectors of the crack $\varGamma$, respectively. Let $+$ and $-$ denote the two sides of $\varGamma$, and for any vector or tensor $\bm{a}$, we let $\bm{a}^\pm$ denote its value evaluated on the $\pm$ side of $\varGamma$. Also we denote $\bm{n}^\pm$ the unit outward normal to the $\pm$ side of $\varGamma$. The jump operator $\llbracket\cdot\rrbracket$ is defined such that 
\begin{equation*}
	\llbracket \bm{a}\cdot\bm{n} \rrbracket = \bm{a}^+ \cdot \bm{n}^+ + \bm{a}^- \cdot \bm{n}^-.
\end{equation*}




Once the solution of problem \eqref{RVE-BVP} is obtained, the total strain energy and the macroscopic stress tensor is given by
\begin{equation}\label{Eq:sigma_micro}
\overline{\bm{\sigma}} = \frac{1}{\meas(\mathcal{B})}\int_{\mathcal{B}_s} \bm{\sigma}\;d\mathcal{B},
\end{equation}
which will be matched with \eqref{Eq:sigma_micro} to obtain the expression of $\overline{\bm{\sigma}}(\overline{\bm{\varepsilon}}, d)$ and $\varPsi(\overline{\bm{\varepsilon}}, d)$, if fitting is needed. Note that $\meas(\mathcal{B})$ denotes the measure of $\mathcal{B}$.

By construction, this RVE behaves differently in tension and in compression, a desired behavior.  
Under the above assumptions, the effective elastic response will be orthotropic in both cases. 
In this work, the numerical solution to the problem \eqref{RVE-BVP} is obtained with the commercial software Abaqus. More details are given in \ref{sec:Abaqus}.
For convenience, we will drop the bar of $\overline{\bm{\sigma}}$ and $\overline{\bm{\varepsilon}}$ from now on within Section \ref{ourmodel}.

\subsection{Ansatz for the macroscopic strain energy density}
From the setup of the RVE, it can be concluded that for a given $d$, if the signs of the components of $\bm{\varepsilon}$ (i.e., the loading mode being tensile or compressive) are fixed, then the secant modulus $\mathbb{C}$ is independent of $\bm{\varepsilon}$. In other words, \emph{$\mathbb{C}$ is also the tangent modulus}. We hereby propose
an ansatz for the macroscopic strain energy density \eqref{psi-macro}:
\begin{equation}\label{Eq:ansatz_strain_energy}
\begin{aligned}
&\varPsi(\bm{\varepsilon},d) =\frac12 \Bigl\{ H(\beta(\bm{\varepsilon})) \;  \underbrace{\bm{\varepsilon}:\mathbb{C}^{t}:\bm{\varepsilon}}_{\text{tension}}+ \left [ 1-H(\beta(\bm{\varepsilon})) \right ] \underbrace{\bm{\varepsilon}:\mathbb{C}^{c}:\bm{\varepsilon}}_{\text{compression}} \Bigr\}\\
&= \frac{1}{2} \Bigl\{H(\beta(\bm{\varepsilon})) \left(\mathbb{C}^{t}_{1111}\varepsilon_{11}^2 + \mathbb{C}^{t}_{2222}\varepsilon_{22}^2 +
\mathbb{C}^{t}_{3333}\varepsilon_{33}^2
+2\mathbb{C}^{t}_{1122}\varepsilon_{11}\varepsilon_{22}\right.\\
&\quad\left.
+2\mathbb{C}^{t}_{1133}\varepsilon_{11}\varepsilon_{33}
+2\mathbb{C}^{t}_{2233}\varepsilon_{22}\varepsilon_{33} +4\mathbb{C}^{t}_{2323}\varepsilon_{23}^{2}
+4\mathbb{C}^{t}_{1313}\varepsilon_{13}^{2}
+4\mathbb{C}^{t}_{1212}\varepsilon_{12}^{2}\right)\\ 
&\quad+ \left[ 1 - H(\beta(\bm{\varepsilon}))\right]\left(\mathbb{C}^{c}_{1111}\varepsilon_{11}^2 + \mathbb{C}^{c}_{2222}\varepsilon_{22}^2 + \mathbb{C}^{c}_{3333}\varepsilon_{33}^2 +2\mathbb{C}^{c}_{1122}\varepsilon_{11}\varepsilon_{22}\right.\\
&\quad\left.
+2\mathbb{C}^{c}_{1133}\varepsilon_{22}\varepsilon_{33}
+2\mathbb{C}^{c}_{2233}\varepsilon_{22}\varepsilon_{33} +4\mathbb{C}^{c}_{2323}\varepsilon_{23}^{2} +4\mathbb{C}^{c}_{1313}\varepsilon_{13}^{2} +4\mathbb{C}^{c}_{1212}\varepsilon_{12}^{2}\right) \Bigr\},
\end{aligned}
\end{equation}
where $H$ is the Heaviside function, and $\beta(\bm{\varepsilon})$, whose expression will be determined later, switches between the cases in which the RVE is in tension or compression, under the given strain $\bm{\varepsilon}$. 

Tensors $\mathbb{C}^{t}=\mathbb{C}^{t}(d)$ and $\mathbb{C}^{c}=\mathbb{C}^{c}(d)$ denote the effective tensile and compressive moduli, respectively, and the sense (tensile vs.~compressive) is determined by $\beta(\bm{\varepsilon})$. 
Here the phase field $d$ is closely related to the crack length ratio $r_a := 2a/L$, where $2a$ is the crack length, whose definition depends on the shape of $\varGamma$, and $L$ is the side length of the RVE. 
For any $d$ (or $r_a$) and $\bm{\varepsilon}$, $\mathbb{C}_{ijkl}^t$ and $\mathbb{C}_{ijkl}^c$ can be obtained either by inspection or by numerically analyzing the RVE. 

Hence we construct $\mathbb{C}_{ijkl}^t$ and $\mathbb{C}_{ijkl}^c$ using the following ansatz 
 \begin{equation*} 
	 \mathbb{C}^{m}_{ijkl}(d) = g(d)\mathbb{P}^{m}_{ijkl} + \mathbb{Q}^{m}_{ijkl}, \quad m=t,c.
 \end{equation*}

Then
\begin{equation}\label{Eq:split_psi}
\begin{split}
\varPsi(\bm{\varepsilon},d) &= \frac12\Bigl\{H(\beta(\bm{\varepsilon})) \left \{  g(d) \; \bm{\varepsilon}:\mathbb{P}^{t}:\bm{\varepsilon} + \bm{\varepsilon}:\mathbb{Q}^{t}:\bm{\varepsilon} \right \} \\
&\quad+ \left [ 1-H(\beta(\bm{\varepsilon})) \right ]\left\{ g(d) \; \bm{\varepsilon}:\mathbb{P}^{c}:\bm{\varepsilon} + \bm{\varepsilon}:\mathbb{Q}^{c}:\bm{\varepsilon}  \right\}\Bigr\}.
\end{split}
\end{equation}

Equations \eqref{Eq:ansatz_strain_energy} and \eqref{Eq:split_psi} then give rise to the following constitutive relations, when the mode $m=t$ or $c$ is known:
\begin{subequations}\label{Eq:sigmas}
\begin{align}
	\sigma_{11} &= \left[g(d)\mathbb{P}_{1111}^m + \mathbb{Q}_{1111}^m\right]\varepsilon_{11} + \left[g(d)\mathbb{P}_{1122}^m + \mathbb{Q}_{1122}^m\right]\varepsilon_{22} +
	\left[g(d)\mathbb{P}_{1133}^m + \mathbb{Q}_{1133}^m\right]\varepsilon_{33}, \label{Eq:sigmas_11}\\
	\sigma_{22} &= \left[g(d)\mathbb{P}_{1122}^m + \mathbb{Q}_{1122}^m\right]\varepsilon_{11} + \left[g(d)\mathbb{P}_{2222}^m + \mathbb{Q}_{2222}^m\right]\varepsilon_{22} + \left[g(d)\mathbb{P}_{2233}^m + \mathbb{Q}_{2233}^m\right]\varepsilon_{33},
	\label{Eq:sigmas_22}\\
	\sigma_{33} &= \left[g(d)\mathbb{P}_{1133}^m + \mathbb{Q}_{1133}^m\right]\varepsilon_{11} + \left[g(d)\mathbb{P}_{2233}^m + \mathbb{Q}_{2233}^m\right]\varepsilon_{22} + \left[g(d)\mathbb{P}_{3333}^m + \mathbb{Q}_{3333}^m\right]\varepsilon_{33},
	\label{Eq:sigmas_33}\\
	\sigma_{23} &= 2\left[g(d)\mathbb{P}_{2323}^m + \mathbb{Q}_{2323}^m\right]\varepsilon_{23},
	\label{Eq:sigmas_23}\\
	\sigma_{13} &= 2\left[g(d)\mathbb{P}_{1313}^m + \mathbb{Q}_{1313}^m\right]\varepsilon_{13},
	\label{Eq:sigmas_13}\\
	\sigma_{12} &= 2\left[g(d)\mathbb{P}_{1212}^m + \mathbb{Q}_{1212}^m\right]\varepsilon_{12}
	\label{Eq:sigmas_12}.
\end{align}
\end{subequations}

\begin{rem}
	Here we do \emph{not} associate tension (or compression) with the degradation function $g(d)$, to make the model more general at the beginning.
\end{rem}

\begin{rem}
	At this point, the remaining task is to determine $\{\mathbb{P}_{ijkl}^m\}$ and $\{\mathbb{Q}_{ijkl}^m\}$, which totals to 36 unknowns. This can be achieved in two means: (a) with prior knowledge, i.e., by making \eqref{Eq:split_psi} reduce to usual expressions in the special cases of uniaxial tension, etc., and (b) by fitting, i.e., by matching \eqref{Eq:sigmas} and \eqref{Eq:sigma_micro}. Our strategy is to first deduce as many coefficients as possible with prior knowledge, and then resort to fitting.
\end{rem}

Before ending this section we list one of the constraints for $\{\mathbb{P}_{ijkl}^m\}$ and $\{\mathbb{Q}_{ijkl}^m\}$, which is that
$\varPsi(\bm{\varepsilon}, d=0) = \varPsi_0(\bm{\varepsilon})$ 
for either the tensile case $H(\beta(\bm{\varepsilon}))=1$ or the compressive case $H(\beta(\bm{\varepsilon}))=0$. That is to say,
\begin{equation}
	\begin{aligned}
	\varPsi_0(\bm{\varepsilon})=&\frac12\Bigl\{\left[\mathbb{P}^m_{1111}+\mathbb{Q}^m_{1111}\right]\varepsilon_{11}^2+\left[\mathbb{P}^m_{2222}+\mathbb{Q}^m_{2222}\right]\varepsilon_{22}^2+\left[\mathbb{P}^m_{3333}+\mathbb{Q}^m_{3333}\right]\varepsilon_{33}^2\\
	&+2\left[\mathbb{P}^m_{1122}+\mathbb{Q}^m_{1122}\right]\varepsilon_{11}\varepsilon_{22}+2\left[\mathbb{P}^m_{1133}+\mathbb{Q}^m_{1133}\right]\varepsilon_{11}\varepsilon_{33}+2\left[\mathbb{P}^m_{2233}+\mathbb{Q}^m_{2233}\right]\varepsilon_{22}\varepsilon_{33}\\
	&+4\left[\mathbb{P}^m_{2323}+\mathbb{Q}^m_{2323}\right]\varepsilon_{23}^2
	+4\left[\mathbb{P}^m_{1313}+\mathbb{Q}^m_{1313}\right]\varepsilon_{13}^2
	+4\left[\mathbb{P}^m_{1212}+\mathbb{Q}^m_{1212}\right]\varepsilon_{12}^2
	\Bigr\},
	\\& m=t,c.
	\end{aligned}
	\label{Eq:psi0_PQm}
\end{equation}

Here for a linear isotropic material, $\varPsi_0(\bm{\varepsilon})$ takes the following form:
\begin{equation}
	\varPsi_0(\bm{\varepsilon}) =  {\frac\lambda{2}}(\varepsilon_{11} + \varepsilon_{22}+ \varepsilon_{33})^2 + \mu\left(\varepsilon_{11}^2 + \varepsilon_{22}^2 + \varepsilon_{33}^2 + 2 \varepsilon_{23}^2+ 2 \varepsilon_{13}^2+ 2 \varepsilon_{12}^2\right).
	\label{Eq:psi0}
\end{equation}

Equating \eqref{Eq:psi0_PQm} and \eqref{Eq:psi0} we get the following equations, for $m = t,c$,
\begin{subequations}\label{Eq:psi0_conditions}
	\begin{align}
	\mathbb{P}^{m}_{1111} + \mathbb{Q}^{m}_{1111} &= 
	\mathbb{P}^{m}_{2222} + \mathbb{Q}^{m}_{2222} = 
	\mathbb{P}^{m}_{3333} + \mathbb{Q}^{m}_{3333} = \lambda + 2\mu, 
	\label{Eq:psi0_conditions_iiii}\\
	\mathbb{P}^{m}_{1122} + \mathbb{Q}^{m}_{1122} &= 
	\mathbb{P}^{m}_{1133} + \mathbb{Q}^{m}_{1133} = 	
	\mathbb{P}^{m}_{2233} + \mathbb{Q}^{m}_{2233} = \lambda, \label{Eq:psi0_conditions_iijj}\\		
	\mathbb{P}^{m}_{2323} + \mathbb{Q}^{m}_{2323} &= 
	\mathbb{P}^{m}_{1313} + \mathbb{Q}^{m}_{1313} = 
	\mathbb{P}^{m}_{1212} + \mathbb{Q}^{m}_{1212} = \mu.
	\label{Eq:psi0_conditions_ijij}
	\end{align}
\end{subequations}

\subsection{Determining most parameters $\{\mathbb{P}^{m}_{ijkl}\}$ and $\{\mathbb{Q}^{m}_{ijkl}\}$ with prior knowlegde}\label{sec:det-most-par}
\vspace{2ex}


In this section we will determine most of the coefficients $\{\mathbb{P}^m_{ijkl}\}$ and $\{\mathbb{Q}^m_{ijkl}\}$ by specializing the model to simple tests.

First, it is noted that the symmetry in the crack plane imply:
\begin{subequations}\label{Eq:sym_Cij}
\begin{align}
\mathbb{P}^m_{1111}=\mathbb{P}^m_{3333}, \quad\mathbb{Q}^m_{1111}=\mathbb{Q}^m_{3333},
\label{Eq:sym_Cij_iiii}\\
\mathbb{P}^m_{1122}=\mathbb{P}^m_{2233}, \quad\mathbb{Q}^m_{1133}=\mathbb{Q}^m_{2233},
\label{Eq:sym_Cij_iijj}\\
\mathbb{P}^m_{1212}=\mathbb{P}^m_{2323}, \quad\mathbb{Q}^m_{1212}=\mathbb{Q}^m_{2323}.
\label{Eq:sym_Cij_ijij}
\end{align}
\end{subequations}


\begin{enumerate}[i.]
\item Tensile tests.
We first consider the uniaxial tension in the $x_2$-direction, i.e., with macroscopic strain $\varepsilon_{22}>0$ while the rest of strain components are zero.
In this special case, the loading is clearly tensile, i.e., $m=t$. Then \eqref{Eq:sigmas_22} yields
\[\sigma_{22}=\left[g(d)\mathbb{P}^{t}_{2222}+\mathbb{Q}_{2222}^t\right]\varepsilon_{22}.\]

We further assume, as consistent with existing models, that the degradation takes full effect in this case, i.e., $\mathbb{Q}^t_{2222} = 0$, which yields $\mathbb{P}^{t}_{2222} = \lambda + 2\mu$ per \eqref{Eq:psi0_conditions_iiii}.

Next we consider the uniaxial tension loading in the $x_1$-direction, i.e., $\varepsilon_{11}>0$ while the other strain components are zero. This is also a tensile case, hence \eqref{Eq:sigmas_22} gives
\[
\sigma_{22}=\left[g(d)\mathbb{P}^{t}_{1122}+\mathbb{Q}_{1122}^t\right]\varepsilon_{11}.
\]

In accordance with numerical results (see Figure \ref{c1122_abaqus}) we found that $\mathbb{Q}^t_{1122} = 0$ and hence \eqref{Eq:psi0_conditions_iijj} can be simplified to
$\mathbb{P}^{t}_{1122} = \lambda$. Also, referring to \eqref{Eq:sym_Cij_iijj}, we obtain $\mathbb{Q}^t_{2233}=0$, and $\mathbb{P}^t_{2233}=\lambda$.

\item Compressive tests.
We now examine the case of the uniaxial compressive test in the $x_2$-direction, i.e., the only non-zero component of strain tensor is $\varepsilon_{22}<0$. Then 
\[\sigma_{22}=\left[g(d)\mathbb{P}^c_{2222}+\mathbb{Q}_{2222}^c\right]\varepsilon_{22}.\]

In this case, there is no degradation, i.e., $\mathbb{P}^c_{2222} = 0$. Equation \eqref{Eq:psi0_conditions_iiii} then gives
 $\mathbb{Q}^{c}_{2222} = \lambda+2\mu$.

We then move on to the case of uniaxial compression in the $x_1$-direction, i.e., $\varepsilon_{11}<0$ while other components are zero. In this case, \eqref{Eq:sigmas_11} and \eqref{Eq:sigmas_22} simplify to
\[
\begin{aligned}
\sigma_{11}&=\left[g(d)\mathbb{P}^c_{1111}+\mathbb{Q}_{1111}^c\right]\varepsilon_{11},\\
\sigma_{22}&=\left[g(d)\mathbb{P}^c_{1122}+\mathbb{Q}_{1122}^c\right]\varepsilon_{11}.
\end{aligned}
\]

Numerical results (see Figures \ref{c1111_abaqus} and \ref{c1122_abaqus}) confirm that there is no degradation in this case, which yields $\mathbb{P}^c_{1111}=\mathbb{P}^c_{1122}=0$. Then, \eqref{Eq:psi0_conditions_iiii} and \eqref{Eq:psi0_conditions_iijj} lead to
$\mathbb{Q}^{c}_{1111} = \lambda+2\mu$ and $\mathbb{Q}^{c}_{1122} = \lambda$, respectively. Also, due to \eqref{Eq:sym_Cij_iiii} and \eqref{Eq:sym_Cij_iijj} we obtain $\mathbb{P}^c_{3333}=\mathbb{P}^c_{2233}=0$, thus $\mathbb{Q}^{c}_{3333} = \lambda+2\mu$ and $\mathbb{Q}^{c}_{2233} = \lambda$.

\item Shear test.

Consider a shear test, with $\varepsilon_{12}\neq0$, while other components are zero.
By symmetry, we can deduce
\begin{equation}\label{Eq:sym_shear_PQm}
	\begin{aligned}
		\mathbb{P}^{t}_{1212} = \mathbb{P}^{c}_{1212}, \quad \mathbb{Q}^{t}_{1212} = \mathbb{Q}^{c}_{1212},\\
		\mathbb{P}^{t}_{1313} = \mathbb{P}^{c}_{1313}, \quad \mathbb{Q}^{t}_{1313} = \mathbb{Q}^{c}_{1313},\\
		\mathbb{P}^{t}_{2323} = \mathbb{P}^{c}_{2323}, \quad \mathbb{Q}^{t}_{2323} = \mathbb{Q}^{c}_{2323}.
	\end{aligned}
\end{equation}
Per \eqref{Eq:sym_shear_PQm}, henceforth we will simply use $\mathbb{P}_{1212}$ and $\mathbb{Q}_{1212}$ to denote $\mathbb{P}_{1212}^m$ and $\mathbb{Q}_{1212}^m$, respectively. 
However, in this case, it is not easy to determine a simple analytic expression to separate $\mathbb{P}_{1212}^m$ from $\mathbb{Q}_{1212}^m$, also $\mathbb{P}_{2323}^m$ from $\mathbb{Q}_{2323}^m$. In the sequel we will resort to fitting to determine these coefficients.

\end{enumerate}

\paragraph{Summary of results so far}
Up to now we have obtained the following results of the coefficients:
\begin{enumerate}[(i)]
	\item $\mathbb{Q}^{t}_{2222}=\mathbb{Q}^{t}_{1122}=\mathbb{Q}^{t}_{2233}=\mathbb{P}^{c}_{1111}=\mathbb{P}^{c}_{3333}=\mathbb{P}^{c}_{2222}=\mathbb{P}^{c}_{1122} = \mathbb{P}^{c}_{2233} = 0$,
	\item $\mathbb{P}^{t}_{2222} = \mathbb{Q}^{c}_{1111} = \mathbb{Q}^{c}_{3333} = \mathbb{Q}^{c}_{2222} =  \lambda+2\mu$,
	\item $\mathbb{P}^{t}_{1122} = \mathbb{P}^{t}_{2233} =\mathbb{Q}^{c}_{1122} =\mathbb{Q}^{c}_{2233} =\lambda$,
	\item $\mathbb{P}_{2323}+\mathbb{Q}_{2323}=\mathbb{P}_{1313}+\mathbb{Q}_{1313}=\mathbb{P}_{1212}+\mathbb{Q}_{1212}=\mu$.
\end{enumerate}

Then the unknown coefficients at this point are:
\begin{equation*}
\begin{aligned}
&\mathbb{P}_{1111}^t,~ \mathbb{Q}_{1111}^t\left(=\lambda+2\mu-\mathbb{P}_{1111}^t\right), \quad 
\mathbb{P}_{1133}^t,~\mathbb{Q}_{1133}^t\left(=\lambda-\mathbb{P}_{1133}^t\right), \\
&\mathbb{P}_{1133}^c,~\mathbb{Q}_{1133}^c\left(=\lambda-\mathbb{P}_{1133}^c\right),\\
&\mathbb{P}_{1212},~ \mathbb{Q}_{1212}(=\mu-\mathbb{P}_{1212}), \quad \mathbb{P}_{1313},~ \mathbb{Q}_{1313}(=\mu-\mathbb{P}_{1313}). 
\end{aligned}
\end{equation*}

With the current results, \eqref{Eq:split_psi} can be simplified as: 
\begin{equation*}
\begin{aligned}
\varPsi(\bm{\varepsilon},d)&=\frac12\lbrace H(\beta(\bm{\varepsilon}))\\&\Bigl\{ g(d)\left[\mathbb{P}^t_{1111}\varepsilon_{11}^2+(\lambda+2\mu)\varepsilon_{22}^2+\mathbb{P}^t_{1111}\varepsilon_{33}^2+2\lambda\varepsilon_{11}\varepsilon_{22}+2\mathbb{P}^t_{1133}\varepsilon_{11}\varepsilon_{33}+2\lambda\varepsilon_{22}\varepsilon_{33}\right]\\ &+(\lambda+2\mu-\mathbb{P}^t_{1111})\varepsilon_{11}^2+(\lambda+2\mu-\mathbb{P}^t_{1111})\varepsilon_{33}^2+2(\lambda-\mathbb{P}^t_{1133})\varepsilon_{11}\varepsilon_{33}\Bigr\}\\&+\left[1-H(\beta(\bm{\varepsilon}))\right]\Bigl\{ g(d)\mathbb{P}^c_{1133}\varepsilon_{11}\varepsilon_{33} \\&+ (\lambda+2\mu)\left(\varepsilon_{11}^2+\varepsilon_{22}^2+\varepsilon_{33}^2\right)+2\lambda\varepsilon_{11}\varepsilon_{22}+2(\lambda-\mathbb{P}^c_{1133})\varepsilon_{11}\varepsilon_{33}+2\lambda\varepsilon_{22}\varepsilon_{33}\Bigr\}\\
&+2g(d)\left(\mathbb{P}_{1212}\varepsilon_{23}^2+\mathbb{P}_{1313}\varepsilon_{13}^2+\mathbb{P}_{1212}\varepsilon_{12}^2\right)+2(\mu-\mathbb{P}_{1212})\left(\varepsilon_{23}^2+\varepsilon_{12}^2\right)+2(\mu-\mathbb{P}_{1313})\varepsilon_{13}^2.
\end{aligned}
\end{equation*}

\subsection{Determining $\mathbb{P}^{t}_{1111}$, $\mathbb{P}^{t}_{1133}$, and $\mathbb{P}^{c}_{1133}$}
According to the numerical experiments (see Figures \ref{c1111_abaqus} and \ref{c1133_abaqus}), the coefficient $\mathbb{P}^{t}_{1111}$ is determined by making 
\[\mathbb{P}^{t}_{1111}\varepsilon_{11}^2 + \left(\lambda+2\mu\right)\varepsilon_{22}^2 + \mathbb{P}^{t}_{1111}\varepsilon_{33}^2 + 2\lambda\varepsilon_{11}\varepsilon_{22} + 2\mathbb{P}_{1133}^t\varepsilon_{11}\varepsilon_{33}+2\lambda\varepsilon_{22}\varepsilon_{33}\]
a perfect square.
This allows us to determine that
\[\mathbb{P}^{t}_{1111} = \mathbb{P}^{t}_{1133} = \frac{\lambda^2}{\lambda + 2\mu},\]
and that $\beta(\bm{\varepsilon})$ should read
\begin{equation*}\label{Eq:beta}
	\beta(\bm{\varepsilon}) = \lambda\varepsilon_{11} + (\lambda+2\mu) \varepsilon_{22}+\lambda\varepsilon_{33},
\end{equation*}
which is just $\sigma_{22}$ for the undamaged material. An alternative expression for $\beta(\bm{\varepsilon})$ is given in terms of the Poisson ratio, $\nu\varepsilon_{11} + (1-\nu)\varepsilon_{22}+\nu\varepsilon_{33}$.

It is also worth mentioning that according to numerical results, we obtain $\mathbb{P}_{1133}^c=0$, and hence $\mathbb{Q}_{1133}^c=\lambda$.

\subsection{Determining $\mathbb{P}_{1212}$ and $\mathbb{P}_{1313}$ by fitting}\label{sec:fitting}
It turns out that $\mathbb{P}_{1212}$ does not seem to depend on the material parameters with a closed-form analytic expression, see Figure \ref{c1122_abaqus}. Hence we fit the non-dimensional expression $\mathbb{P}_{1212}/\mu$ as a function of the phase field $d$, for various Poisson ratios.
More precisely, we fit $\mathbb{P}_{1212}/\mu$ as a quadratic function of $d$ for $d\in [0, 1]$, with the coefficients dependent on $\nu$.


Here one constraint follows. When $d=1$, $\varepsilon_{12}\neq0$, $\varepsilon_{11}=\varepsilon_{22}=\varepsilon_{33}=0$, we need to enforce $\sigma_{12}=0$, otherwise the through-crack shear test would not go through, see Section \ref{sec:through-crack}. Thus we need $\mathbb{P}_{1212}=\mu$ when $d=1$.
With this constraint, we write \[\mathbb{P}_{1212}=\mu[a(\nu)(1-d)^2 + b(\nu)(1-d) + 1].\]  

The coefficients $a(\nu)$ and $b(\nu)$ obtained from least square fits are given in Table \ref{relationships}. Note that for $d$ close to 1, an analysis based on \cite{MARKENSCOFF20121478} may be able to yield more accurate expression. 

\begin{table}[htbp]
	\centering
	\captionsetup{justification=raggedright,singlelinecheck=false}
	\caption{Fitted coefficients for $\mathbb{P}_{1212}=\mu[a(\nu)(1-d)^2 + b(\nu)(1-d) + 1]$. Each row is fitted by least squares with four data points.}
	\begingroup\setlength{\fboxsep}{0pt}
	\colorbox{lightgray}{%
		\begin{tabular}{ccccc}
			\toprule
			$\nu$ & $a(\nu)$ & $b(\nu)$  & Correlation coefficients\\
			\midrule
			$0.2$ & $-0.0728$ & $-0.5896$ & 0.9999\\
			\midrule
			$0.25$ & $-0.0042$ & $-0.7022$ & 0.9999\\
			\midrule
			$0.3$ & $0.1281$ & $-0.8783$ & 1.0000\\
			\midrule
			$0.35$ & $0.2807$ & $-1.0886$ & 1.0000\\
			\midrule
			$0.4$ & $0.3952$ & $-1.2633$ & 0.9999\\
			\midrule
			$0.45$ & $1.0790$ & $-1.9825$ & 0.9994\\
			\bottomrule
		\end{tabular}
	}\endgroup
	\label{relationships}
\end{table}

Also, according to the numerical results, we obtain:
	\[\mathbb{P}_{1313}=0,\quad \mathbb{Q}_{1313}=\mu.\]

\subsection{Summary of the proposed phase field model}

In summary, the strain energy density $\varPsi$ is given as follows:
\begin{equation}\label{Eq:psi_final}
\begin{split}
\varPsi(\bm{\varepsilon},d) = 
H(\beta(\bm{\varepsilon}))\varPsi_t 
+ \bigl(1 - H(\beta(\bm{\varepsilon}))\bigr)\varPsi_c 
+ \varPsi_s.
\end{split}
\end{equation}

We then provide the expressions of relevant quantities in different variables.
\paragraph{The model expressed with the $x_2$-axis normal to the crack $\varGamma$}
In this case,
\begin{equation}\label{special-orientation}
\begin{aligned}
\beta(\bm{\varepsilon}) &= \lambda \varepsilon_{11} + \left(\lambda+2\mu\right) \varepsilon_{22} + \lambda\varepsilon_{33},\\
\varPsi_t&:= \frac1{2(\lambda+2\mu)}
\left[g\left(d\right) \left(\lambda\varepsilon_{11}+(\lambda+2\mu)\varepsilon_{22}+\lambda\varepsilon_{33}\right)^2 + 4\mu(\lambda+\mu)\left(\varepsilon_{11}^2+\varepsilon_{33}^2\right)+4\lambda\mu\varepsilon_{11}\varepsilon_{33}\right],\\ 
\varPsi_c&:=\frac12\left [ \lambda\left(\varepsilon_{11}+\varepsilon_{22}+\varepsilon_{33}\right)^2+2\mu\left(\varepsilon_{11}^2+\varepsilon_{22}^2+\varepsilon_{33}^2\right)\right],\\
\varPsi_s&:=g_s(d;\nu)2\mu\left(\varepsilon_{23}^2+\varepsilon_{12}^2\right)+2\mu\varepsilon_{13}^2,
\end{aligned}
\end{equation}
where 
\[g_s(d;\nu)=1+[g(d)-1]\left(a(\nu)(1-d)^2 + b(\nu)(1-d) + 1 \right)\]
is named the \emph{shear degradation function}.

\paragraph{The model expressed in terms of invariants and pseudo-invariants of the strain tensor}
To obtain an expression suitable for any crack orientation, we define two invariants
\begin{equation*}
\begin{aligned}
I_1&=\tr\bm{\varepsilon}=\varepsilon_{11}+\varepsilon_{22}+\varepsilon_{33},\\
I_2&=\frac12\left[\left(\tr\bm{\varepsilon}\right)^2-\tr\left({\bm{\varepsilon}}^2\right)\right]=
\varepsilon_{11}\varepsilon_{22}+\varepsilon_{11}\varepsilon_{33}+\varepsilon_{22}\varepsilon_{33}-\varepsilon_{23}^2-\varepsilon_{13}^2-\varepsilon_{12}^2,
\end{aligned}
\end{equation*}
and two pseudo-invariants
\[I_4=\bm{n}\cdot\bm{\varepsilon}\bm{n}, \quad I_5=\bm{n}\cdot\bm{\varepsilon}^{2}\bm{n},\] where \[\bm{n}:=\frac{\nabla d}{\|\nabla d\|}.\] 

Now we can express the terms of \eqref{Eq:psi_final} in terms of these invariants and pseudo-invariants as:
\begin{equation}\label{with-invariants}
\begin{aligned}
\beta(\bm{\varepsilon}) &:= \lambda I_1  + 2\mu I_4,
\\
\varPsi_t&:= \frac1{2(\lambda+2\mu)}
\left[g(d) \left(\lambda I_1 + 2\mu I_4\right)^2 + 4\mu(\lambda+\mu)\left(I_1^2+I_4^2-2I_2-2I_5\right)+4\lambda\mu\left(I_2-I_1I_4+I_5\right)\right],
\\
\varPsi_c&:=\frac12\left [ \lambda I_1^2 + 2\mu\left(I_1^2+2I_4^2-2I_2-2I_5\right)\right],
\\
\varPsi_s&:=g_s(d;\nu)2\mu \left(I_5-I_4^2\right).
\end{aligned}
\end{equation}

%
%

The derivation is given in \ref{sec:invariants}.

	With the above development, in \emph{any} coordinate system, i.e., not necessarily related to the crack orientation, the constitutive relation can be obtained as, in Voigt's notation,
\begin{equation*}
\begin{aligned}
	&\bm{\sigma}(\bm{\varepsilon},d) = \frac{\partial\varPsi}{\partial\bm{\varepsilon}} = \frac{\partial\varPsi}{\partial I_1} 
\begin{Bmatrix}
1 \\ 1 \\ 1 \\ 0 \\ 0 \\ 0
\end{Bmatrix}
+ \frac{\partial\varPsi}{\partial I_2}
\begin{Bmatrix}
\varepsilon_{22} + \varepsilon_{33} \\ \varepsilon_{11} + \varepsilon_{33} \\ \varepsilon_{11} + \varepsilon_{22} \\ -\varepsilon_{23} \\ -\varepsilon_{13} \\ -\varepsilon_{12}
\end{Bmatrix}
+ \frac{\partial\varPsi}{\partial I_4}
\begin{Bmatrix}
n_1^2 \\ n_2^2 \\ n_3^2 \\ n_2n_3 \\ n_1n_3 \\ n_1n_2
\end{Bmatrix}\\
&\quad+ \frac{\partial\varPsi}{\partial I_5}
\begin{Bmatrix}
2\varepsilon_{11}n_1^2 + 2\varepsilon_{12}n_1n_2 + 2\varepsilon_{13}n_1n_3\\ 2\varepsilon_{21}n_1n_2 + 2\varepsilon_{22}n_2^2 + 2\varepsilon_{23}n_2n_3\\ 2\varepsilon_{31}n_1n_3 + 2\varepsilon_{32}n_2n_3 + 2\varepsilon_{33}n_3^2\\
\varepsilon_{31}n_1n_2 + \varepsilon_{32}n_2^2 + \varepsilon_{33}n_2n_3 + \varepsilon_{21}n_1n_3 + \varepsilon_{22}n_2n_3 + \varepsilon_{23}n_3^2\\
\varepsilon_{31}n_1^2 + \varepsilon_{32}n_1n_2 + \varepsilon_{33}n_1n_3 + \varepsilon_{11}n_1n_3 + \varepsilon_{12}n_2n_3 + \varepsilon_{13}n_3^2\\
\varepsilon_{21}n_1^2 + \varepsilon_{22}n_1n_2 + \varepsilon_{23}n_1n_3 + \varepsilon_{11}n_1n_2 + \varepsilon_{12}n_2^2 + \varepsilon_{13}n_2n_3\\
\end{Bmatrix}.
\end{aligned}
\end{equation*}
	
	The corresponding moduli can be written as
	\[
	\begin{aligned}
	\mathbb{C}(\bm{\varepsilon},d) = \frac{\partial^2\varPsi}{\partial I_1^2} \begin{Bmatrix} 1 \\ 1 \\ 1 \\ 0 \\ 0 \\ 0 \end{Bmatrix} \begin{Bmatrix} 
	1 \\ 1 \\ 1 \\ 0 \\ 0 \\ 0 \end{Bmatrix}^T
	+ \frac{\partial\varPsi}{\partial I_2} \begin{bmatrix}
	0 & 1 & 1 & 0 & 0 & 0 \\ 1 & 0 & 1 & 0 & 0 & 0\\ 1 & 1 & 0 & 0 & 0 & 0 \\ 0 & 0 & 0 & -\frac12 & 0 & 0 \\ 0 & 0 & 0 & 0 & -\frac12 & 0 \\ 0 & 0 & 0 & 0 & 0 & -\frac12
	\end{bmatrix}
	+ \frac{\partial^2\varPsi}{\partial I_4^2} \begin{Bmatrix} n_1^2 \\ n_2^2 \\ n_3^2 \\ n_2n_3 \\ n_1n_3 \\ n_1n_2 \end{Bmatrix} \begin{Bmatrix} n_1^2 \\ n_2^2 \\ n_3^2 \\ n_2n_3 \\ n_1n_3 \\ n_1n_2 \end{Bmatrix}^T\\
	+ \frac{\partial\varPsi}{\partial I_5} 
	\begin{bmatrix}
	2n_1^2 & 0 & 0 & 0 & n_1n_3 & n_1n_2 \\ 0 & 2n_2^2 & 0 & n_2n_3 & 0 & n_1n_2\\ 0 & 0 & 2n_3^2 & n_2n_3 & n_1n_3 & 0 \\ 0 & n_2n_3 & n_2n_3 & \frac{n_3^2+n_2^2}{2} & \frac{n_1n_2}{2} & \frac{n_1n_3}{2} \\ n_1n_3 & 0 & n_1n_3 & \frac{n_1n_2}{2} & \frac{n_1^2+n_3^2}{2} & \frac{n_2n_3}{2} \\ n_1n_2 & n_1n_2 & 0 & \frac{n_1n_3}{2} & \frac{n_2n_3}{2} & \frac{n_1^2+n_2^2}{2}
	\end{bmatrix}.
	\end{aligned}
	\]
	
	As now $\varPsi$ depends on $\nabla d$, the strong form for $d$ is modified to be
	\[
		\frac{\partial\varPsi}{\partial d}-\nabla\cdot\frac{\partial \varPsi}{\partial(\nabla d)} + g_c\left( \frac{d}{\ell} - \ell \Delta d\right) = 0\quad \text{in } \Omega,
	\]
	where
	\[
		\begin{aligned}
			\nabla\cdot\frac{\partial\varPsi}{\partial(\nabla d)} = H(\beta(\bm{\varepsilon}))\left(\frac{\partial \varPsi_t}{\partial I_4}\nabla\cdot\frac{\partial I_4}{\partial(\nabla d)}+\frac{\partial \varPsi_t}{\partial I_5}\nabla\cdot\frac{\partial I_5}{\partial(\nabla d)}\right) \\+ \frac{\partial \varPsi_s}{\partial I_4}\nabla\cdot\frac{\partial I_4}{\partial(\nabla d)}+\frac{\partial \varPsi_s}{\partial I_5}\nabla\cdot\frac{\partial I_5}{\partial(\nabla d)},
		\end{aligned}
	\]
where the partial derivatives can be obtained via
\[\frac{\partial I_4}{\partial\bm{n}} = 2\bm{\varepsilon}\bm{n},\quad
\frac{\partial I_5}{\partial\bm{n}} = 2\bm{\varepsilon}^2\bm{n},\]
and
\[
\frac{\partial\bm{n}}{\partial(\nabla d)} = \frac{1}{\|\nabla d\|}\bm{1} - \frac{1}{\|\nabla d\|^3} \nabla d \otimes \nabla d.
\]

Finally, in order to regularize the abrupt change of the crack normal where $\nabla d$ is near zero, we introduce a replacement of $I_4$ and $I_5$ in the constitutive expressions in the form of
\begin{equation*}
\hat{I}_{4,5} = \tanh\left(\alpha l^2 |\nabla d|^2\right)I_{4,5},
\end{equation*}
where $\alpha$ is a small positive number. Here we take $\alpha=5.66\times 10^{-4}$. 

\section{Numerical examples}
\label{numerical}
In this section, the proposed micromechanics-informed phase field model is compared with a few existing models using several numerical examples.
These examples are categorized into two types. Sections \ref{sec:uniaxial} through 
\ref{sec:circular} solve problems in the macroscopic scale, including a uniaxial tension test, a compression test, a shear test, a symmetric bending test, and two through-crack tests. In these examples the proposed model is compared against earlier models, i.e., the isotropic, V-D, and spectral models.
These results show that the proposed model gives reasonable crack paths in all cases. 
Sections \ref{sec:biaxial} and \ref{sec:3d_shaer} compare the constitutive response for a fixed phase field value among a number of models. In these examples, the proposed model is compared with the SK model and the SS models, as well as the earlier models. We will see that only the proposed model show a very close response to the benchmark, by construction.

In the sequel, the marker \texttt{Isotropic} refers to the original model proposed by \cite{bourdin2000numerical}, \texttt{V-D} the volumetric-deviatoric decomposition model proposed by \cite{amor2009regularized}, \texttt{Spectral} the spectral decomposition model proposed by \cite{miehe2010phase}, \texttt{SK} the model proposed by \cite{Steinke2018}, and \texttt{SS1} and \texttt{SS2} the two models proposed by \cite{Strobl2015155}. Finally, \texttt{Proposed} refers to the proposed micromechanics-informed phase field model.

Here we use an in-house code for the macroscopic simulations, while for getting the benchmark solution with the RVE analysis in Sections \ref{sec:biaxial} and \ref{sec:3d_shaer}, the software Abaqus is adopted.


\subsection{Uniaxial tension test}\label{sec:uniaxial}
The first example is a benchmark simulation of a uniaxial tension test. Consider a square plate with a horizontal initial crack located at the middle height from the left side to the center of the specimen.

The material constants are chosen as in Table \ref{tension_material_parameters}.
\begin{table}[htbp]
	\centering
	\captionsetup{justification=raggedright,singlelinecheck=false}
	\caption{Material parameters used in the tension and compression simulations}
	\begingroup\setlength{\fboxsep}{0pt}
	\colorbox{lightgray}{%
		\begin{tabular}{cccccc}
			\toprule
			 $\lambda$ (GPa) & $\mu$ (GPa) & $g_{c}\; \mathrm{(mJ/mm^2)}$ & $l $ (mm)\\
			\midrule
			 $121.15$ & $80.77$ & 2.7 & 40 \\
			\bottomrule
		\end{tabular}
	}\endgroup
	\label{tension_material_parameters}
\end{table}

\begin{figure}[htbp!]
	\centering
	\includegraphics[width=\linewidth]{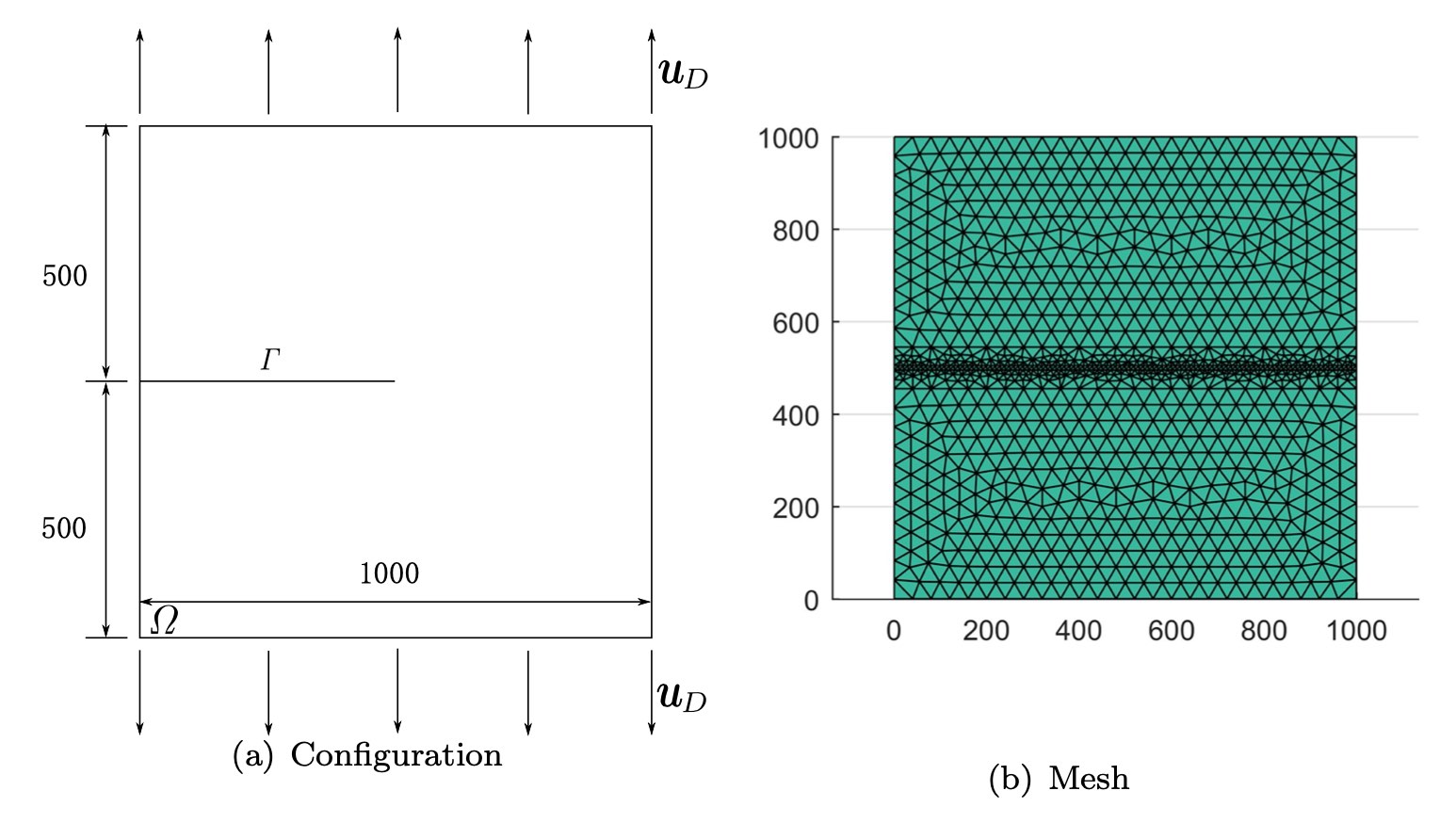}
	\caption{Schematic and mesh of the uniaxial tension test [unit: mm].}
	\label{uniaxialtension}
\end{figure}

The boundary value problem and the mesh are illustrated in Figure \ref{uniaxialtension}. The domain is $\varOmega=[0, 1000]^2$ and the pre-existing crack is $\varGamma=(0,500)\times\{500\}$. We set $d=1$ at $\varGamma$, leave the left and right edges traction free, and minimize \eqref{variational} with $\bm{u}_D\equiv u_D \bm{e}_2$ on the top edge $[0,1000]\times\{1000\}$ and $\bm{u}_D= -u_D \bm{e}_2$ on the bottom edge $[0,1000]\times\{0\}$. For each load step we update the boundary condition as $u_D \leftarrow u_D + \Delta u$, where the displacement increment $\Delta u$ is set to  $0.01$mm, until when $u_D = 0.07$mm, after which the increment is reduced to $\Delta u = 0.0005$mm. The simulation is then run until the sample is totally broken. 


\begin{figure}[htbp!]
	\centering
	\includegraphics[width=\linewidth]{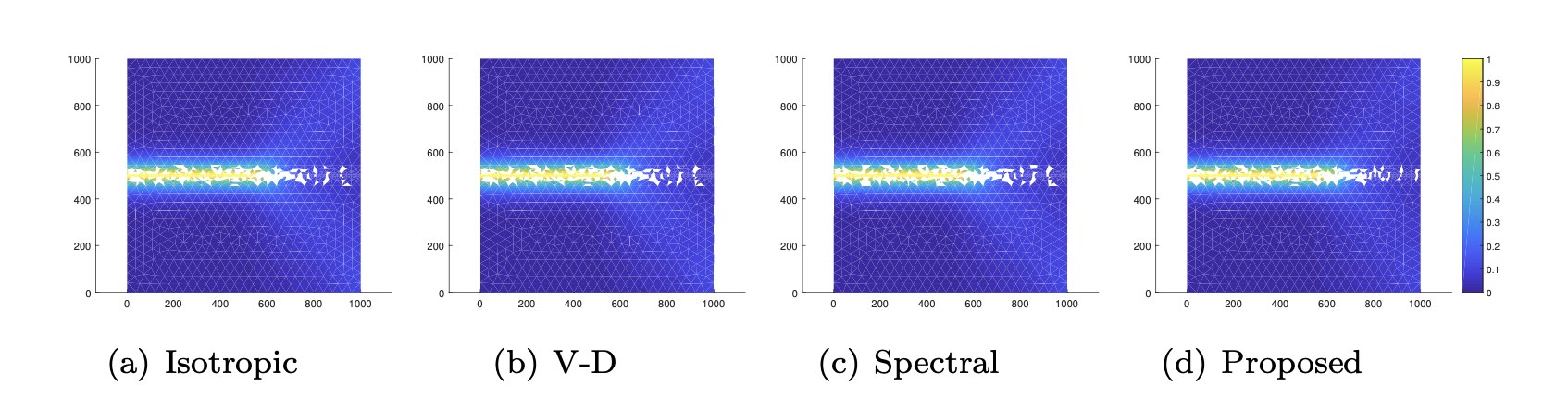}
	\caption{Phase field profiles obtained from the uniaxial tension test at $u_D = 0.080$mm.}
	\label{tension008}
\end{figure}

\begin{figure}[htbp!]
	\centering
	\includegraphics[width=\linewidth]{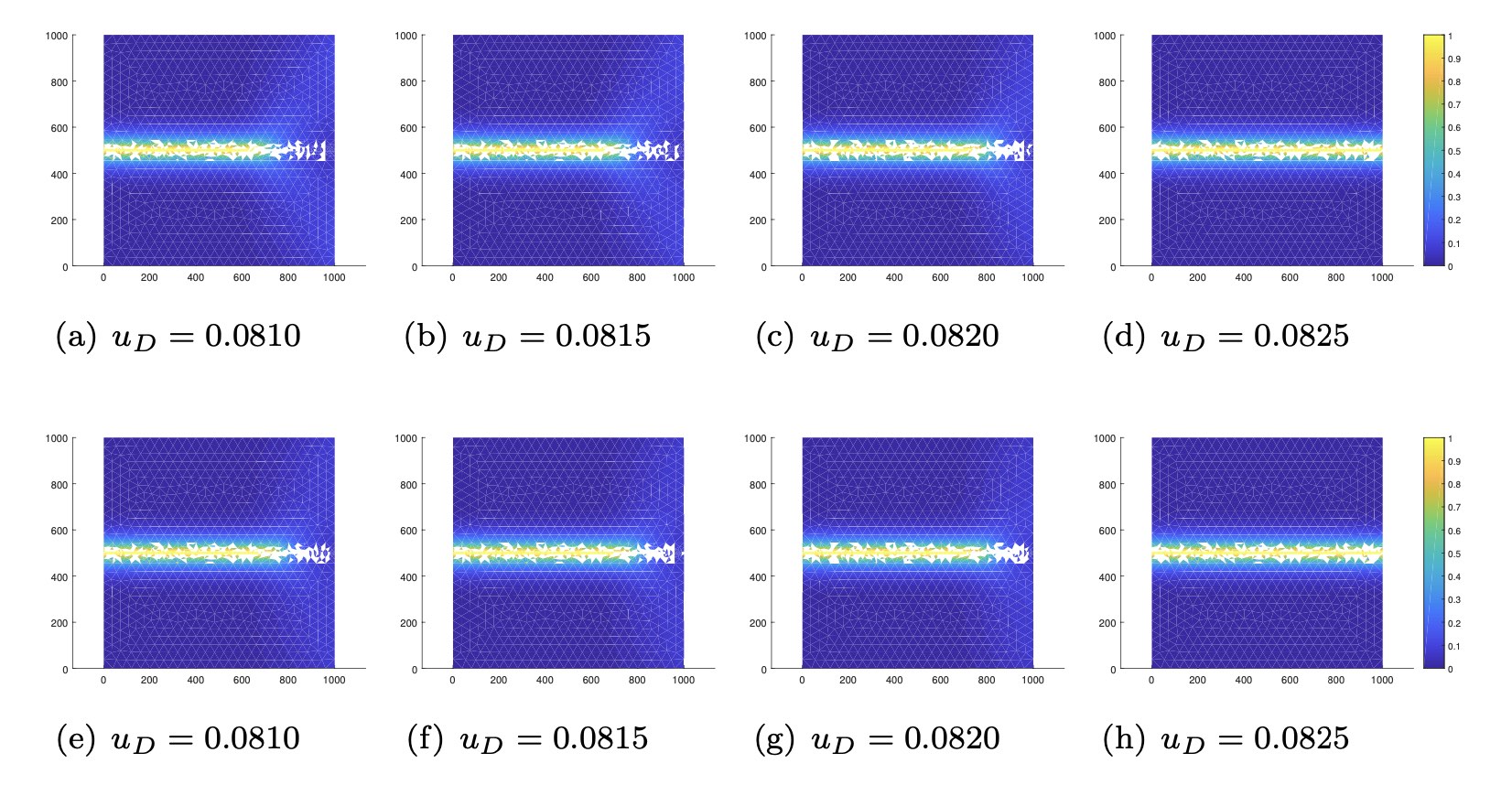}
	\caption{Phase field profiles of the uniaxial tension test from the V-D [(a)-(d)] and proposed [(e)-(h)] models at various load levels.}
	\label{TModelB-N}
\end{figure}

Figure \ref{tension008} compares the phase field profiles among different models at a certain displacement load. This load is prior to the one corresponding to the rapid propagation of the crack.
From Figure \ref{tension008} it can be concluded that the crack starts to propagate almost at the same displacement load. In addition, the results of the isotropic, V-D, and spectral models are almost the same, while the crack in the proposed model propagates a little bit faster. Next, in Figure \ref{TModelB-N}, the results of  the V-D model and the proposed model are compared.


After increasing the displacement continuously, it can be found that the crack in the V-D model propagate faster, while the crack in the proposed model propagates slower, and finally both of them will be totally broken when the displacement reaches $u_D = 0.0825$mm. 

\begin{figure}[htbp]
	\centering
	\includegraphics[width=\linewidth]{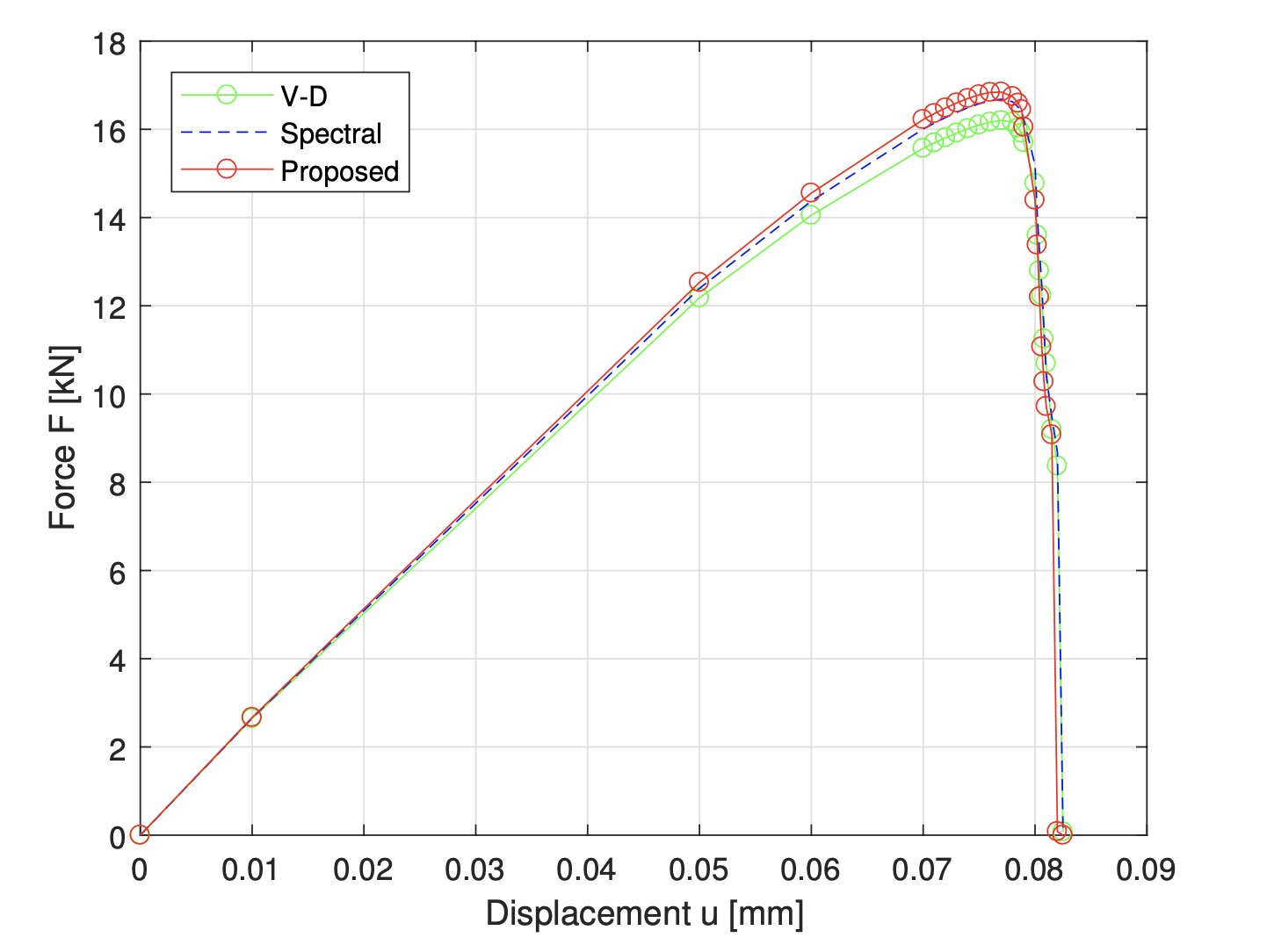}
 	\caption{Force-displacement curves of the uniaxial tension test resulting from the V-D, spectral and proposed phase field models.}
	\label{tension_curve}
\end{figure}

The force-displacement curves of the three models are shown in the Figure \ref{tension_curve}. Overall speaking, the compared models behavior very similarly for a uniaxial tension load.

\subsection{Uniaxial compression test}
The second example is a simulation to show the behaviors of the models under uniaxial compression.
The material constants are given in Table \ref{tension_material_parameters}.
			
The mesh given in Figure \ref{uniaxialtension}(b) is used. The only difference is that the displacement condition is opposite. As mentioned previously, the isotropic model is not expected to distinguish between tension and compression while the others are. 

\begin{figure}[htbp]
	\centering
	\includegraphics[width=\linewidth]{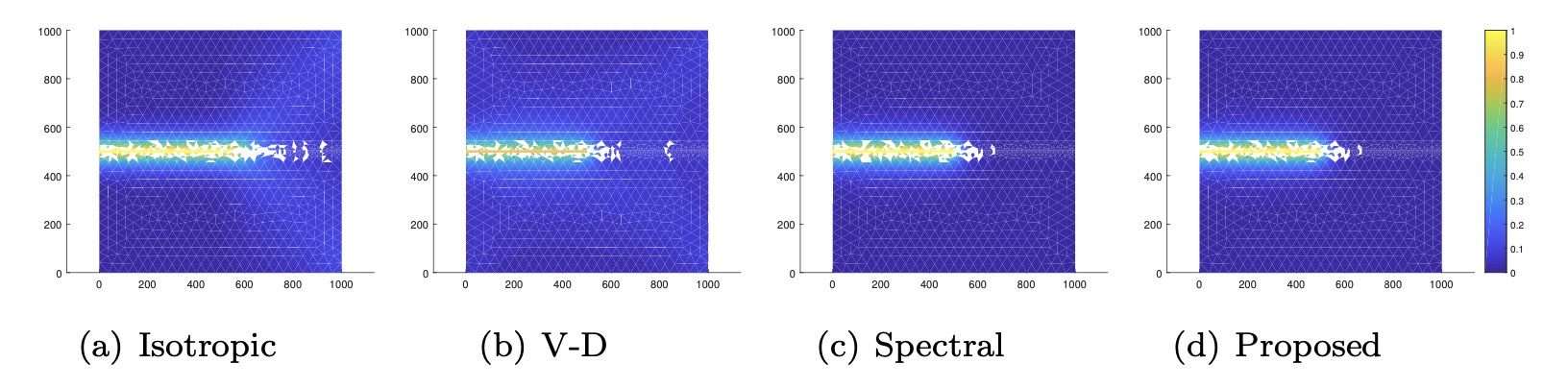}
	\caption{Phase field profiles obtained from the uniaxial compression test at $u_D = 0.08$mm.}
	\label{compression008}
\end{figure}

The the phase field profiles resulting from different models at $u_D=0.08$mm are shown in the Figure \ref{compression008}. From Figure \ref{compression008}, it can be seen that when the displacement increases, the crack in the isotropic model will propagate, as is expected from the setup of this model. On the other hand, although the crack in the V-D model does not propagate, the phase field value increases in some area. This is because in a uniaxial loading, the deviatoric part of the strain is not zero. In contrast, in the spectral and proposed models, the phase field value does not increase all over the domain. 



\begin{figure}[htbp]
	\centering
	\includegraphics[width=\linewidth]{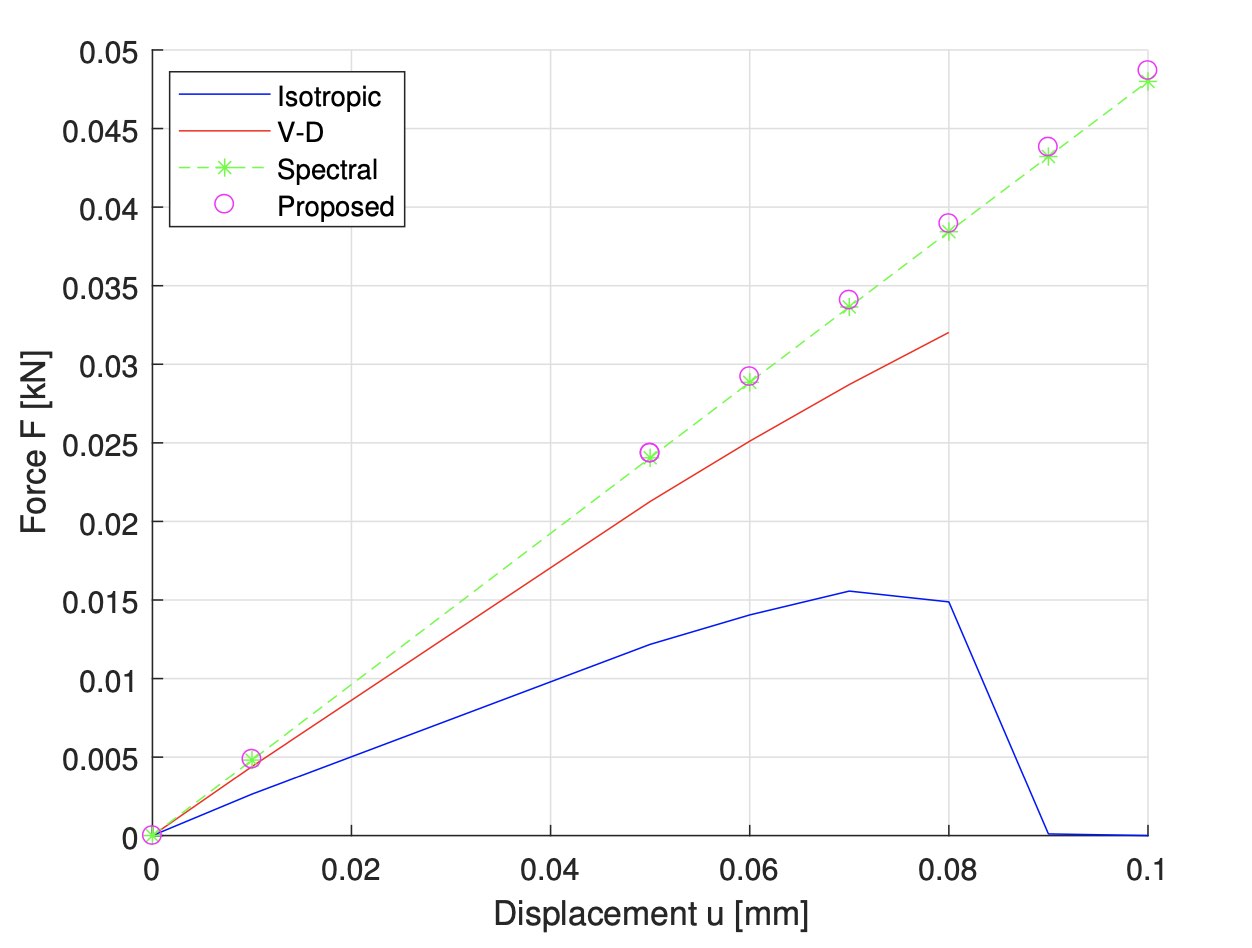}
	\caption{Force-displacement curves of the uniaxial compression test resulting from different models.}
	\label{compression_curve}
\end{figure}

The force-displacement curves are plotted in Figure \ref{compression_curve}. The isotropic model shows the least stiffness, and the (unphysical) fully broken state in which the force decreases to 0 can be clearly observed. 
The spectral and proposed models display almost the same stiffness, while the V-D model is less stiff, indicating that some part of the solid is degraded, in accordance with the observable phase field profile in Figure \ref{compression008}(b), which shows a slight damage. Again, this is attributed to the non-zero deviatoric strain under a uniaxial compression load. 

\begin{figure}[htbp]
	\centering
	\includegraphics[width=\linewidth]{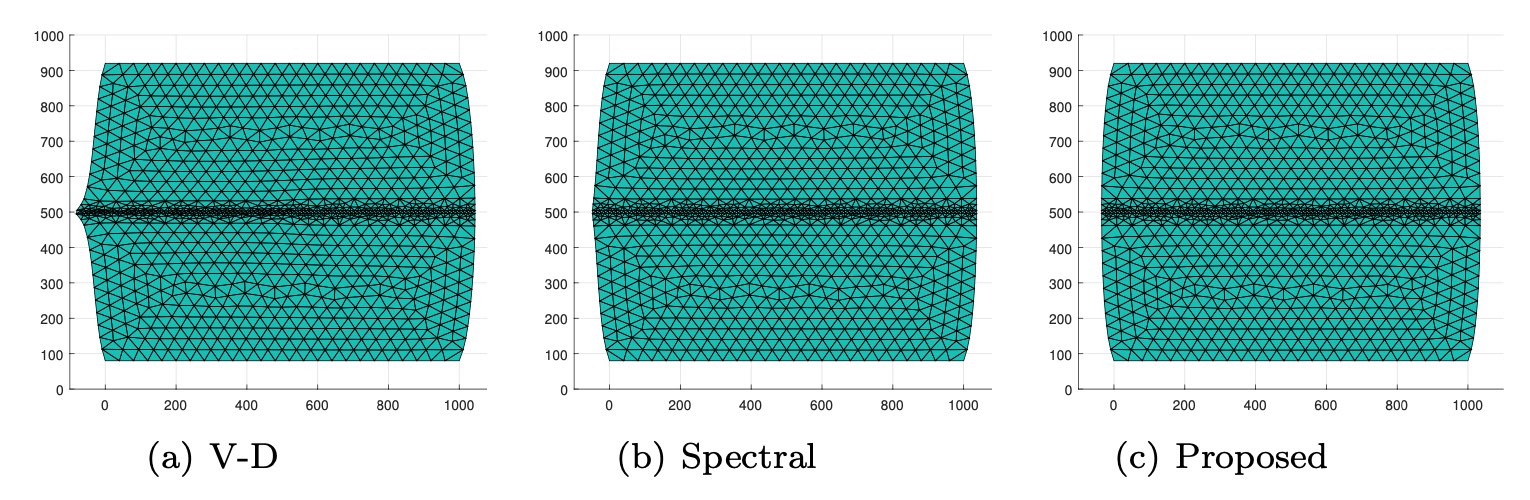}
	\caption{Deformed mesh (with the displacement field 1000 times magnified) of the uniaxial compression test at $u_D=0.08$mm.}
	\label{compressionstep6}
\end{figure}

The deformed configurations at $u_D=0.08$mm of three of the models with the displacement field magnified are shown in Figure \ref{compressionstep6}. It can be seen that the deformed mesh of the V-D model seems anomalous, agreeing with the anomalous behaviors compared with the other two models. On the contrary, the deformed meshes of the spectral and proposed models show a mirror symmetry with respect to the line $x_1=500$, as if the crack does not exist. Based on the above results, we can conclude that within the load range, only the spectral and proposed models perform reasonably in the uniaxial compression test.


 \subsection{Three-point bending test}
This test is a classical benchmark problem which has been analyzed in, for example, \cite{miehe2010phase}.
The material constants of this test are shown in Table \ref{bending_material_parameters}, which are the same as those used in \cite{miehe2010phase}.

\begin{table}[htbp]
	\centering
	\captionsetup{justification=raggedright,singlelinecheck=false}
	\caption{Material parameters used in the three-point bending simulations}
	\begingroup\setlength{\fboxsep}{0pt}
	\colorbox{lightgray}{%
		\begin{tabular}{cccccc}
			\toprule
			$\lambda$(GPa) & $\mu$(GPa) & $g_{c} \mathrm{(mJ/mm^2)}$ & $l $(mm)\\
			\midrule
			$8$ & $12$ & 0.5 & 0.06 \\
			\bottomrule
		\end{tabular}
	}\endgroup
	\label{bending_material_parameters}
\end{table}

\begin{figure}[htbp]
	\centering
	\includegraphics[width=\linewidth]{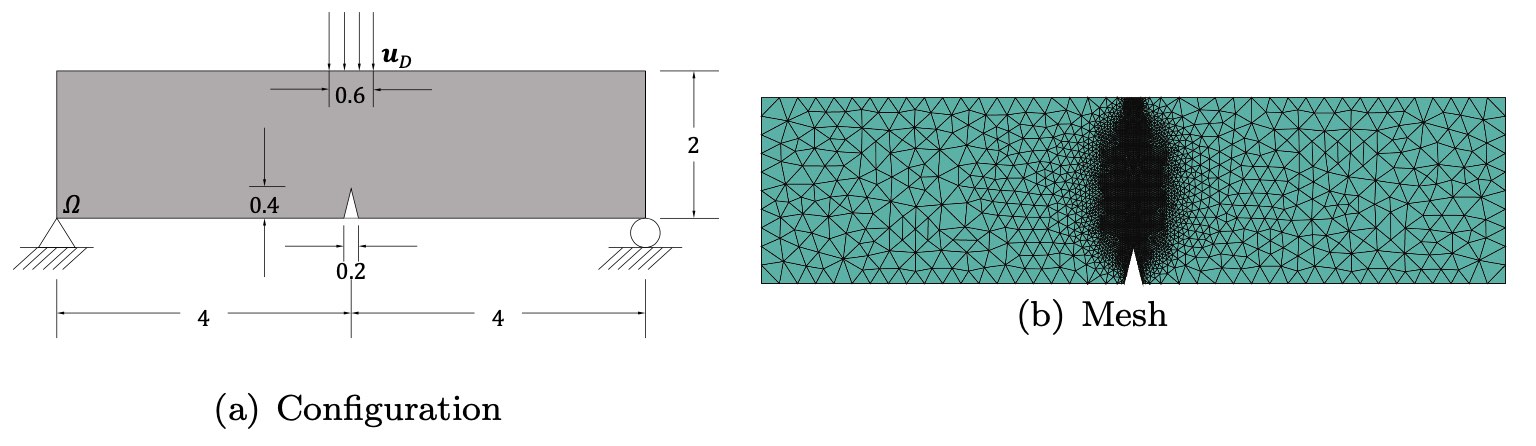}
	\caption{Configuration and mesh of the three-point bending [unit: mm]}
	\label{threepointbending}
\end{figure}

The geometry, loads, and the mesh are illustrated in Figure \ref{threepointbending}.  The domain $\varOmega$ is the shaded part, which is contained in the box $(0,8)\times(0,2)$. We minimize \eqref{variational} with $\bm{u}_D= -u_D\bm{e}_2$ on the central of top edge $[3.7, 4.3]\times\{2\}$, where $u_D \leftarrow u_D + \Delta u$. The displacement increment $\Delta u$ is set to  $0.01$mm when $u_D \in [0,0.04) \cup [0.05,0.12]$mm. In order to closely track the crack propagation, the increment is reduced to $\Delta u = 0.002$mm when $u_D\in [0.04,0.05)$mm. The effective mesh size is also adopted as in \cite{miehe2010phase}.

\begin{figure}[htbp]
	\centering
	\includegraphics[width=\linewidth]{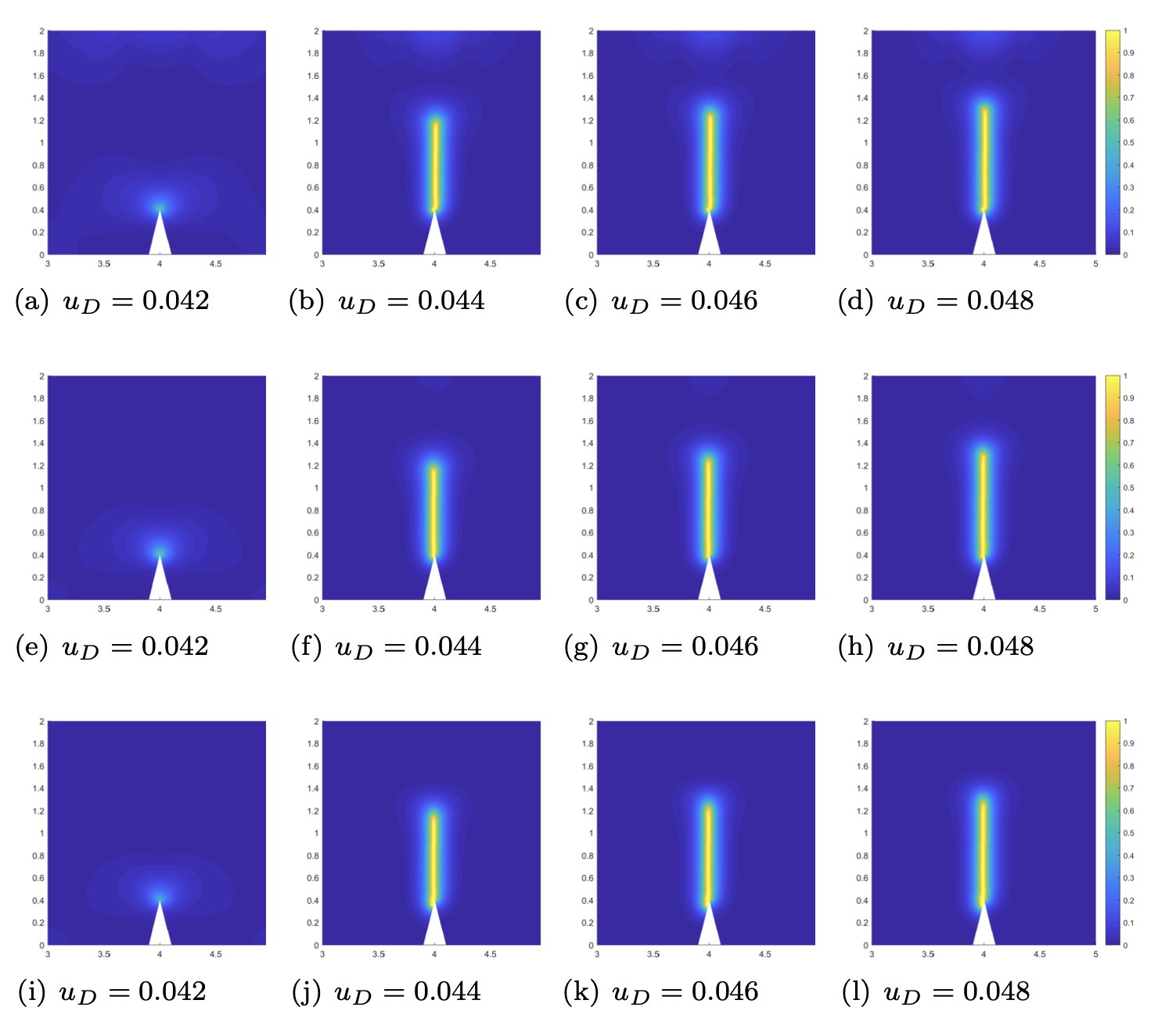}
	\caption{Phase field profiles of the three-point bending test from the V-D [(a)-(d)], spectral [(e)-(h)], and proposed models[(i)-(l)] for $u_D$ in the range of $[0.042,0.048]$.}
	\label{tpb1}
\end{figure}

\begin{figure}[htbp]
	\centering
	\includegraphics[width=\linewidth]{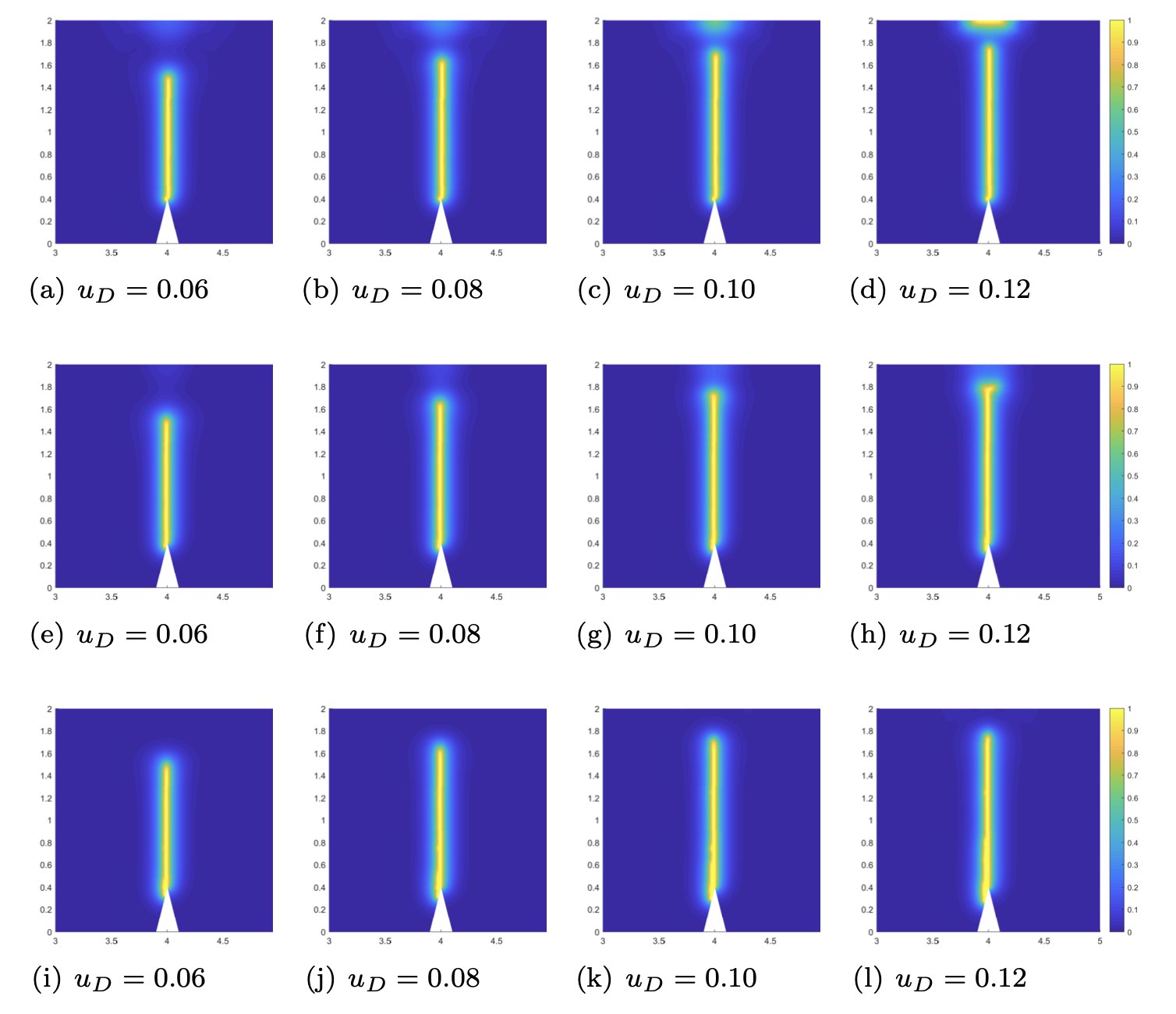}
	\caption{Phase field profiles of the three-point bending test from the V-D [(a)-(d)], spectral [(e)-(h)] and proposed models [(i)-(l)] for $u_D$ in the range of $[0.06, 0.12]$.}
	\label{tpb2}
\end{figure}

We show the crack evolution for the V-D, spectral and proposed models in Figures \ref{tpb1} and \ref{tpb2}.  
Figure \ref{tpb1}(a)(e)(i) show the phase field profiles from different models at the beginning of crack propagation and we observe some initial phase field at equilibrium around the notch. Figure \ref{tpb1}[(a)-(d)] and Figure \ref{tpb2}[(a)-(d)] illustrate the phase field profiles of the bending tests resulting from the V-D model. It can be observed that the phase field of the element around the point where the displacement is applied (the compression area) will increase. This is due to the non-zero deviatoric strain in a uniaxial compression state, and this was also reported in \cite{ziaei2016massive}. On the other hand, Figures \ref{tpb1} and \ref{tpb2} show the phase field results from the spectral model and the proposed model, respectively, both of which are reasonable. And from Figure \ref{tpb2}, we can see that when $u_D=0.12$ mm, the plate is almost fully broken for both cases.

\begin{figure}[htbp]
	\centering
	\includegraphics[width=\linewidth]{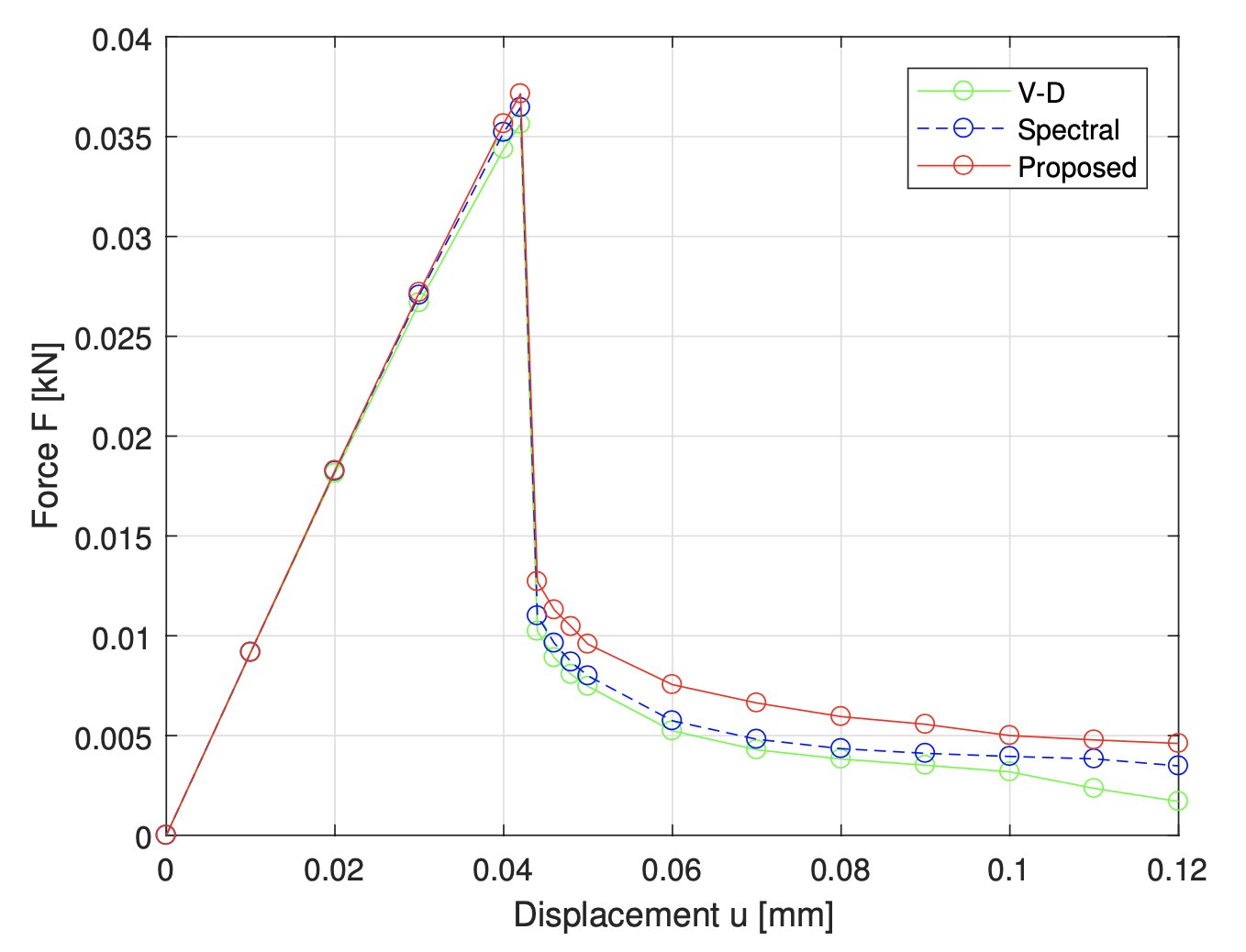}
	\caption{Force-displacement curves of the three-point bending test for three models.}
	\label{bending_curve}
\end{figure}

Figure \ref{bending_curve} shows the force-displacement curves resulting from the simulations for the V-D, spectral, and proposed models. The results conform to what we observed in the phase field profiles. 

It can be concluded that only the spectral and proposed models give acceptable results for the three-point bending test.

\subsection{Shear test}
Now we investigate a square plate of a different size for a shear test.
The material constants are chosen as shown in Table \ref{shear_material_parameters}.

\begin{table}[htbp]
	\centering
	\captionsetup{justification=raggedright,singlelinecheck=false}
	\caption{Material parameters used in the shear simulations}
	\begingroup\setlength{\fboxsep}{0pt}
	\colorbox{lightgray}{%
		\begin{tabular}{cccccc}
			\toprule
			 $\lambda$(GPa) & $\mu$(GPa) & $g_{c} \mathrm{(mJ/mm^2)}$ & $l $(mm)\\
			\midrule
			 $121.15$ & $80.77$ & 2.7 & 3.125 \\
			\bottomrule
		\end{tabular}
	}\endgroup
	\label{shear_material_parameters}
\end{table}

\begin{figure}[htbp]
	\centering
	\includegraphics[width=\linewidth]{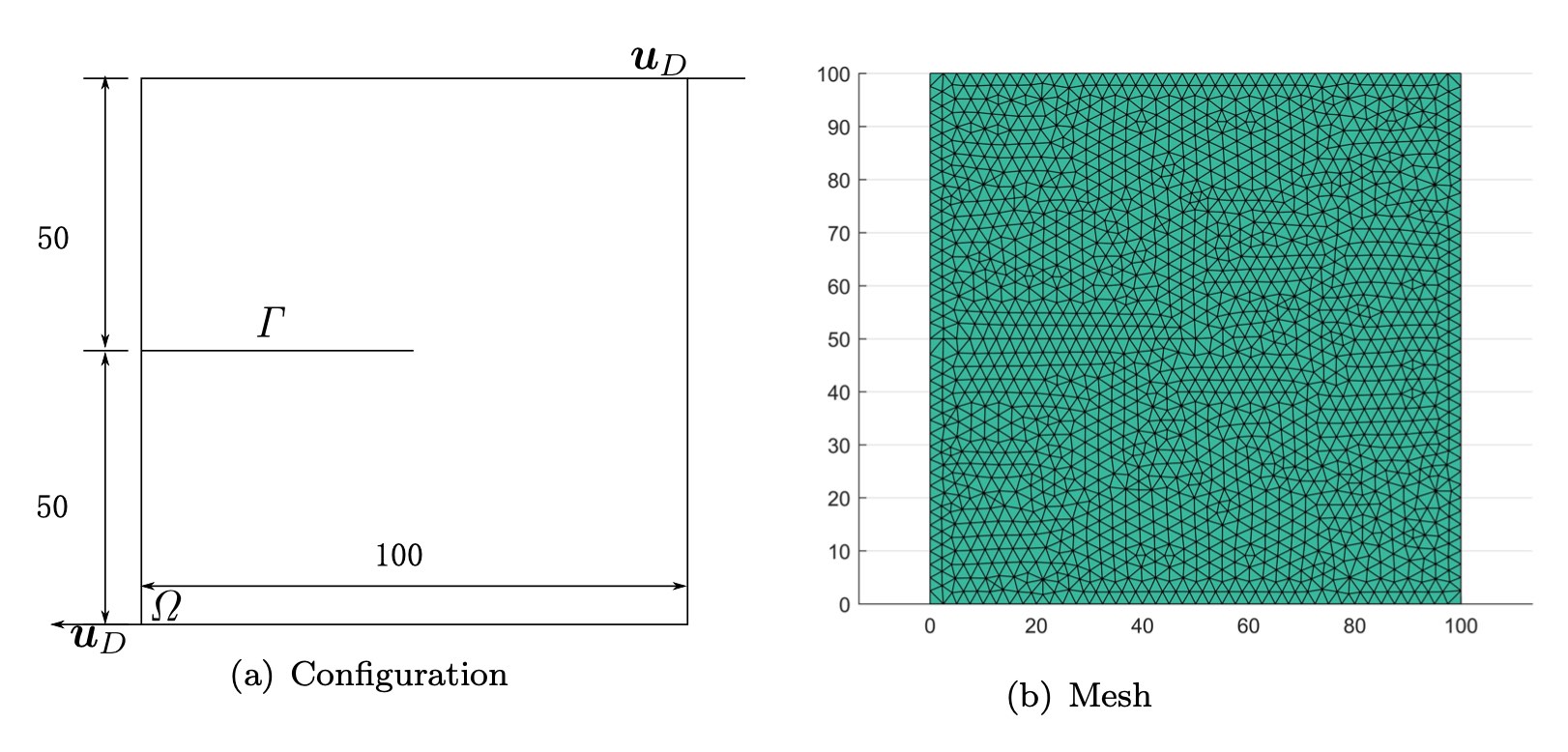}
	\caption{Configuration and mesh of the shear test [unit: mm].}
	\label{sheartest}
\end{figure}

The boundary value problem and the mesh are illustrated in Figure \ref{sheartest}. The domain has a pre-existing crack $\varGamma=(0,50)\times\{50\}$, and $\varOmega=[0,100]^2\setminus\varGamma$, as shown in Figure \ref{sheartest}(a). We minimize \eqref{variational} with $\bm{u}_D\equiv u_D\bm{e}_1$ on the top edge $[0,100]\times\{100\}$, $\bm{u}_D\equiv -u_D\bm{e}_1$ on the bottom edge $[0,100]\times\{0\}$, $u_D\leftarrow u_D + \Delta u$ and the left and right edges and $\varGamma$ traction free. The displacement increment $\Delta u$ is set to  $0.01$mm when $u_D \in [0,0.05) \cup [0.06,0.1]$mm. In order to closely follow the crack propagation, the increment is reduced to $\Delta u = 0.001$mm when $u_D\in [0.05,0.06)$mm.
%

The crack propagation paths predicted by different models are quite different in this test.   In this case, we will only compare the results between V-D, the spectral and the proposed models since the isotropic model does not distinguish between tension and compression, resulting in unphysical branching crack paths, which has been reported by Ziaei-Rad and Shen \citep{ziaei2016massive}.

\begin{figure}[htbp]
	\centering
	\includegraphics[width=\linewidth]{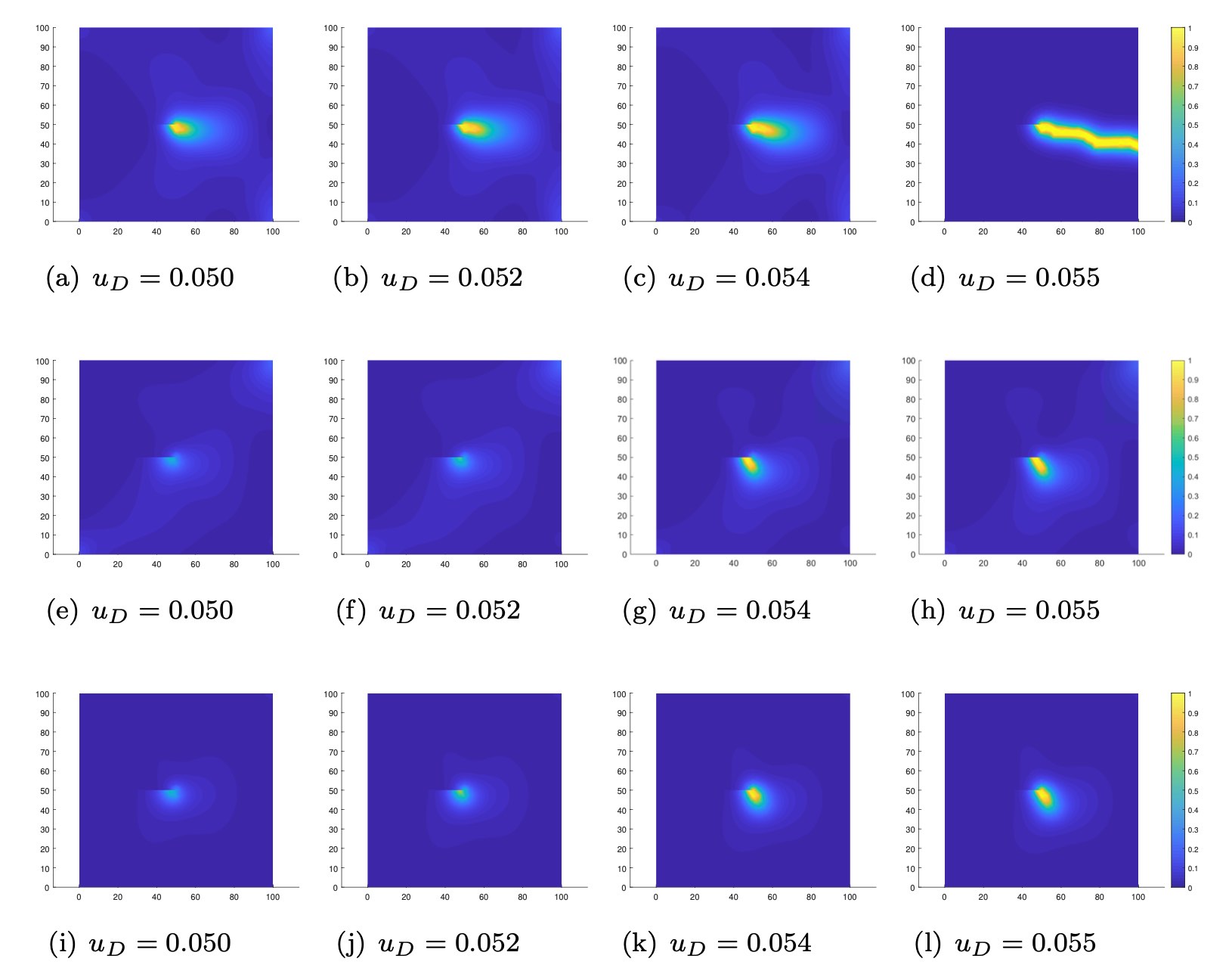}
	\caption{Phase field profiles of the shear test from the V-D [(a)-(d)], spectral [(e)-(h)] and proposed [(i)-(l)] models till $u_D = 0.055$mm.}
	\label{ShearBCN}
\end{figure}

\begin{figure}[htbp]
	\centering
	\includegraphics[width=\linewidth]{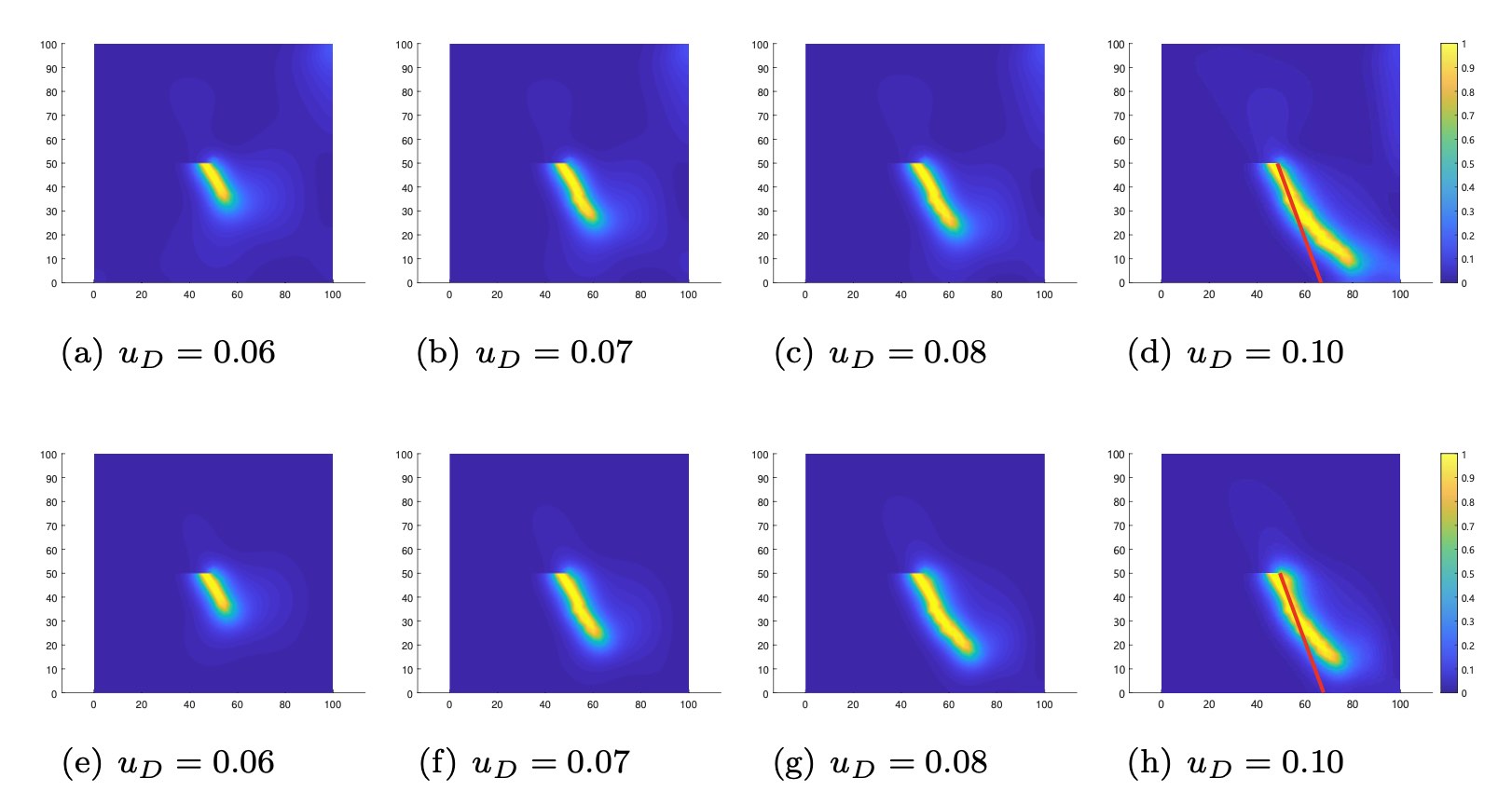}
	\caption{Phase field profiles of the shear test from the spectral [(a)-(d)] and proposed [(e)-(h)] models till $u_D = 0.10$mm. The red lines in (d) and (h) show the initial crack deflection angle predicted from classical linear elastic fracture mechanics. See the text for more details.}
	\label{ShearCN}
\end{figure}

 Figure \ref{ShearBCN} shows the phase field profiles of the shear test from V-D, the spectral and proposed model for $u_D\in[0.050,0.055]$mm. It is shown that the crack pattern in V-D model only has a small angle between horizontal line, while the crack patterns in the spectral and the proposed model are more similar and have a much larger angle. Also, the crack in V-D model run through the material at $u_D = 0.055$mm. The result is similar to that of \cite{ziaei2016massive}. It can also be seen that the displacement load in the spectral model necessary for crack propagation is much larger than the counterpart in the V-D model, which means a stiffer material response by the spectral model. \cite{nguyen2012imposing} also reported this phenomenon. Figure \ref{ShearCN} shows the subsequent crack propagation of the spectral and proposed models.

\begin{figure}[!htbp]
	\centering
	\includegraphics[width=\linewidth]{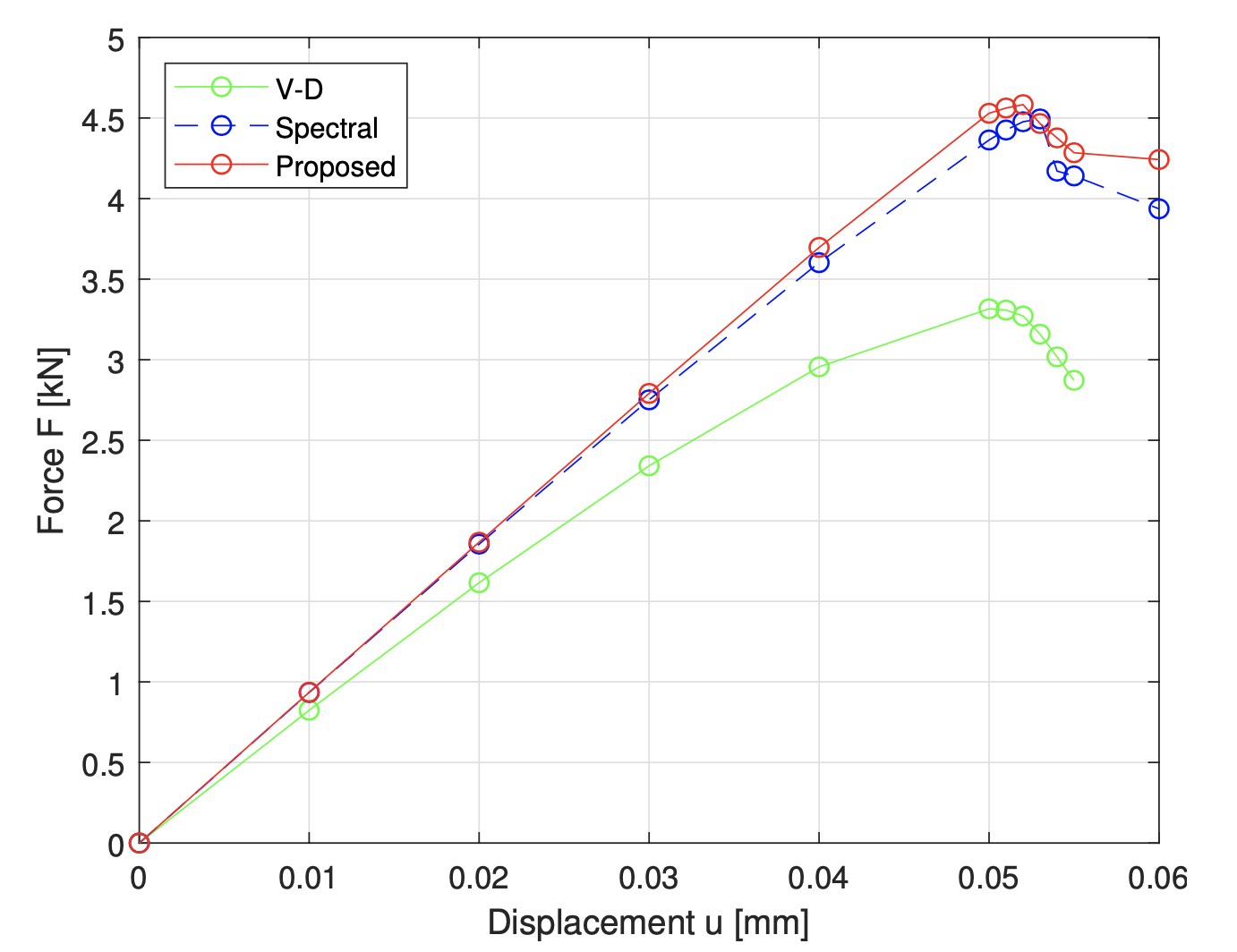}
	\caption{Force-displacement curves of the shear test resulting from different models.}
	\label{shear_curve}
\end{figure}

The red lines in Figure \ref{ShearCN}(d)(h) show the initial crack paths given by the classical mixed-mode loading result from the maximum circumferential stress criterion \citep{erdogan1963}, which gives the crack extension direction according to
 \begin{equation*}
	 \theta_c = 2\arctan\left\{{1\over 4}\left[\frac{K_I}{K_{II}} \pm \sqrt{\left(\frac{K_I}{K_{II}}\right)^2+8}\right]\right\},
 \end{equation*}
where $\theta_c$ is the deviation angle from the extension line of the current crack. For this case the stress intensity factors $K_I=0$, $K_{II}>0$, hence $\theta_c=2\arctan(\sqrt{8}/4)\approx 73^\circ$.
 
Figure \ref{shear_curve} shows the force-displacement curves for the shear test. Again, in this test, the V-D model differs from the spectral and proposed models. 

 \subsection{Through-crack shear test}\label{sec:through-crack}
One of the criticisms [see \cite{strobl2016constitutive}] that the spectral model \citep{miehe2010phase} receives is its inability to output a reasonable result in the through-crack shear test. Hence here we will subject the proposed model and other models to this test. Consider a square plate with a horizontal through crack which is placed at the middle height of the specimen, then a shear load is applied at the top and bottom of the plate. As the crack is through, the domain is actually made of two independent parts with a smooth interface. Hence the correct response is that the two parts slide relative to each other. Note that our model is designed to perform reasonably by construction, see Section \ref{sec:fitting}.

\begin{figure}[!htbp]
 	\centering
 	\includegraphics[width=\linewidth]{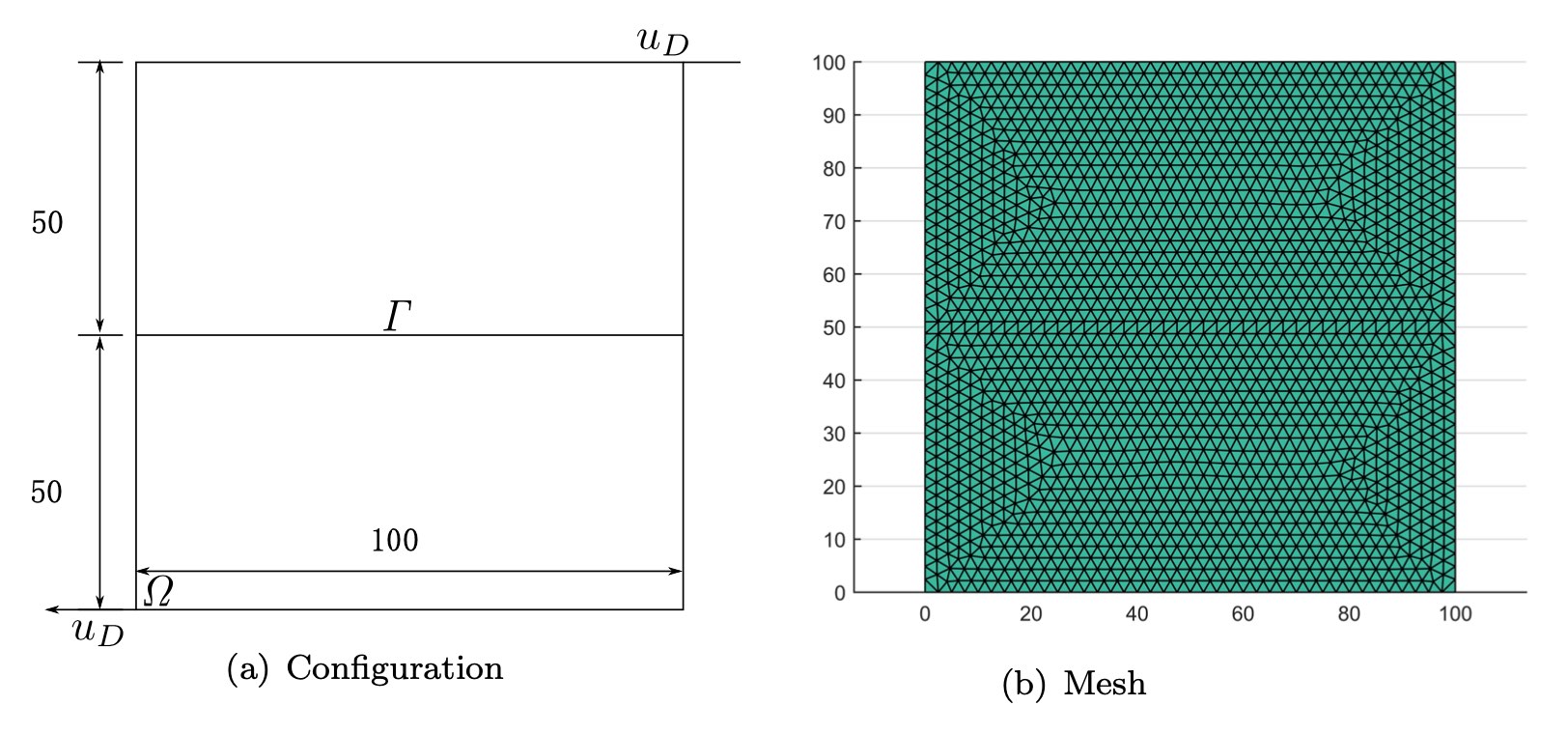}
 	\caption{Configuration and mesh for the two through-crack tests [unit: mm].}
 	\label{through_crack_configuration}
 \end{figure}

  The material constants for the simulations are provided in Table \ref{through_shear_material_parameters}.
 The configuration and mesh are shown in Figure \ref{through_crack_configuration}. The domain is $\varOmega = [0,100]^2$, and the pre-existing crack $\varGamma= [0,100]\times\{50\}$ is described by setting the phase field $d=1$ on $\varGamma$. We minimize \eqref{variational} with $\bm{u}_D = u_D\bm{e}_1$ on the top edge $[0,100]\times\{100\}$, $\bm{u}_D= - u_D\bm{e}_1$ on the bottom edge $[0,100]\times\{0\}$ where $u_D = 0.01$mm, and the remaining portions of $\partial\varOmega$ traction free.
 
  \begin{table}[htbp]
 	\centering
 	\captionsetup{justification=raggedright,singlelinecheck=false}
 	\caption{Material parameters used in the through-crack shear simulations.}
 	\begingroup\setlength{\fboxsep}{0pt}
 	\colorbox{lightgray}{%
 		\begin{tabular}{cccccc}
 			\toprule
 			$\lambda$(GPa) & $\mu$ (GPa) & $g_{c} \mathrm{(mJ/mm^2)}$ & $l $(mm)\\
 			\midrule
 			$121.15$ & $80.77$ & 2.7 & 3.125 \\
 			\bottomrule
 		\end{tabular}
 	}\endgroup
 	\label{through_shear_material_parameters}
 \end{table}


\begin{figure}[!htbp]
 	\centering
 	\includegraphics[width=\linewidth]{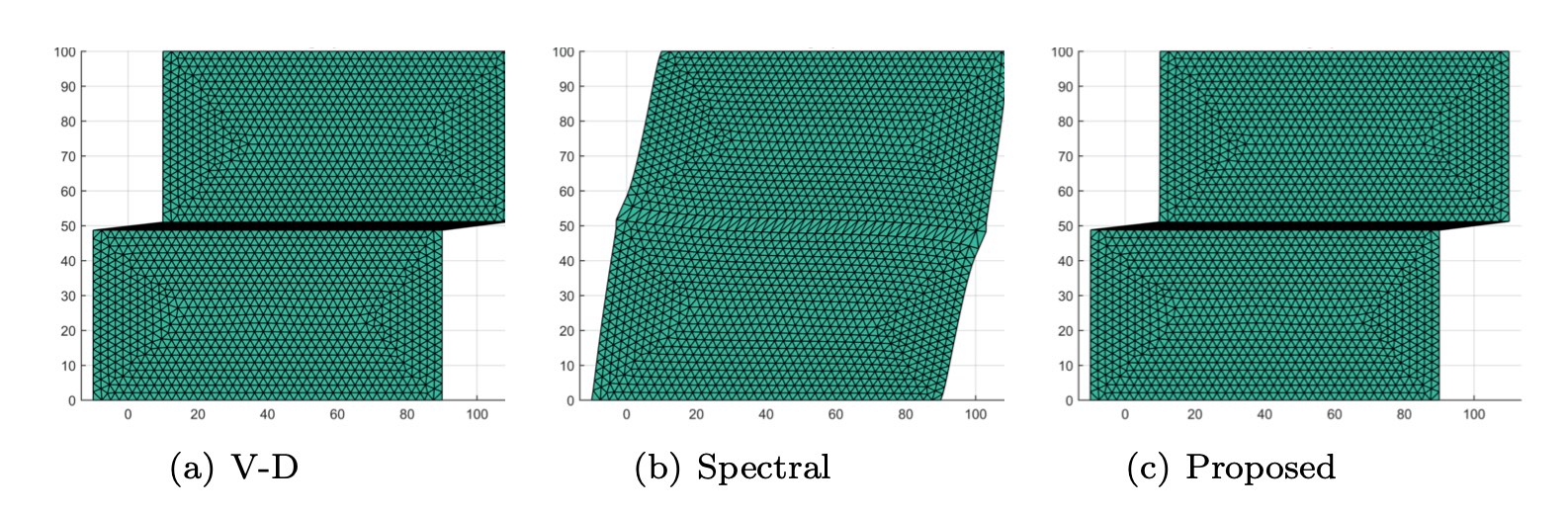}
 	\caption{Deformed mesh (with the displacement field 1000 times magnified) of the through-crack shear test for different models.}
 	\label{through_crack_deformed}
 \end{figure}

Figure \ref{through_crack_deformed} shows the deformed configuration of the V-D, spectral, and proposed models, with the displacement field magnified.
 The results from the V-D and spectral models are consistent to those in \cite{strobl2016constitutive}. In contrast, the spectral model shows an erroneous response, which can be seen by setting $\varepsilon_{11}=\varepsilon_{22}=0$ and $d=1$, which still yields a non-zero $\sigma_{12}$, resulting in some remnant stiffness. It can be concluded that only the V-D and the proposed models give reasonable result for the through-crack shear test among the three compared.
 
\subsection{Through-crack test with a circular load path}
\label{sec:circular}

In order to make sure the result of the through-crack test is valid for different load combinations, we will subject the proposed model and other models to the following test. Using the configuration and mesh shown in Figure \ref{through_crack_configuration}, we fix the bottom of the sample, leave the left and right sides traction free, and impose the following Dirichlet boundary condition on the top edge:
 \begin{equation}\label{circular}
	\bm{u}_{D} = \Delta u \sin \theta \; \bm{e}_1 + \Delta u(1 + \cos \theta) \;\bm{e}_2,\quad \theta \in \{0, \pi/4, \pi/2, 3\pi/4, \pi\},
 \end{equation}
where $\Delta u = 0.01$mm.

\begin{figure}[!htbp]
	\centering
	\includegraphics[width=\linewidth]{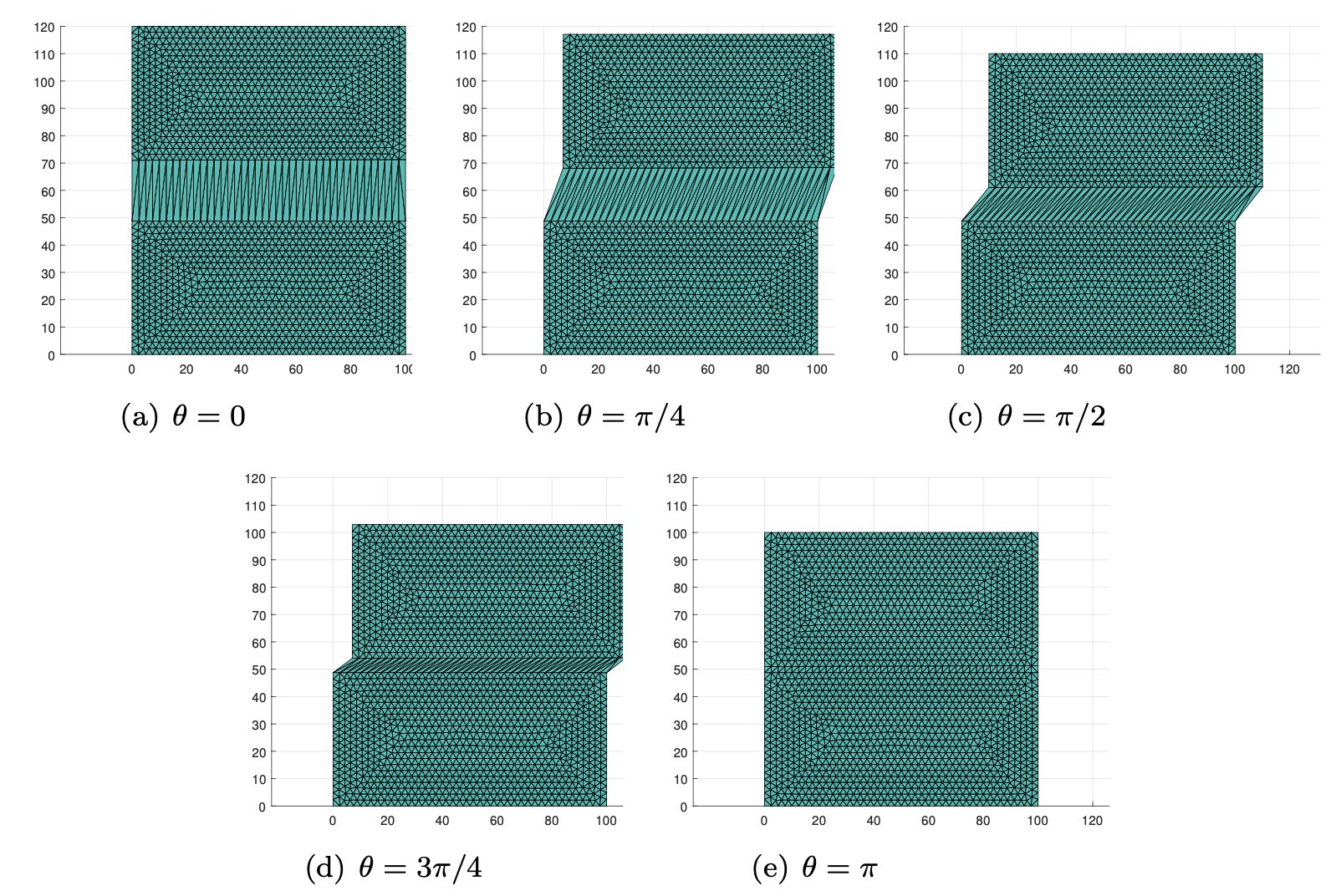}
	\caption{Deformed configurations for the proposed model (with the displacement field 1000 times magnified) of the through-crack test with a circular load path \eqref{circular}.}
	\label{through_crack_n}
\end{figure}

Figure \ref{through_crack_n} illustrates the deformed configuration (with the displacement field magnified) of the proposed models for different $\theta$'s. The results of the V-D model are very similar, and are omitted for the sake of brevity.
 

\subsection{Constitutive response: 2D shear test}
\label{sec:biaxial}
\begin{figure}[!htbp]
	\centering
	\includegraphics[width=\linewidth]{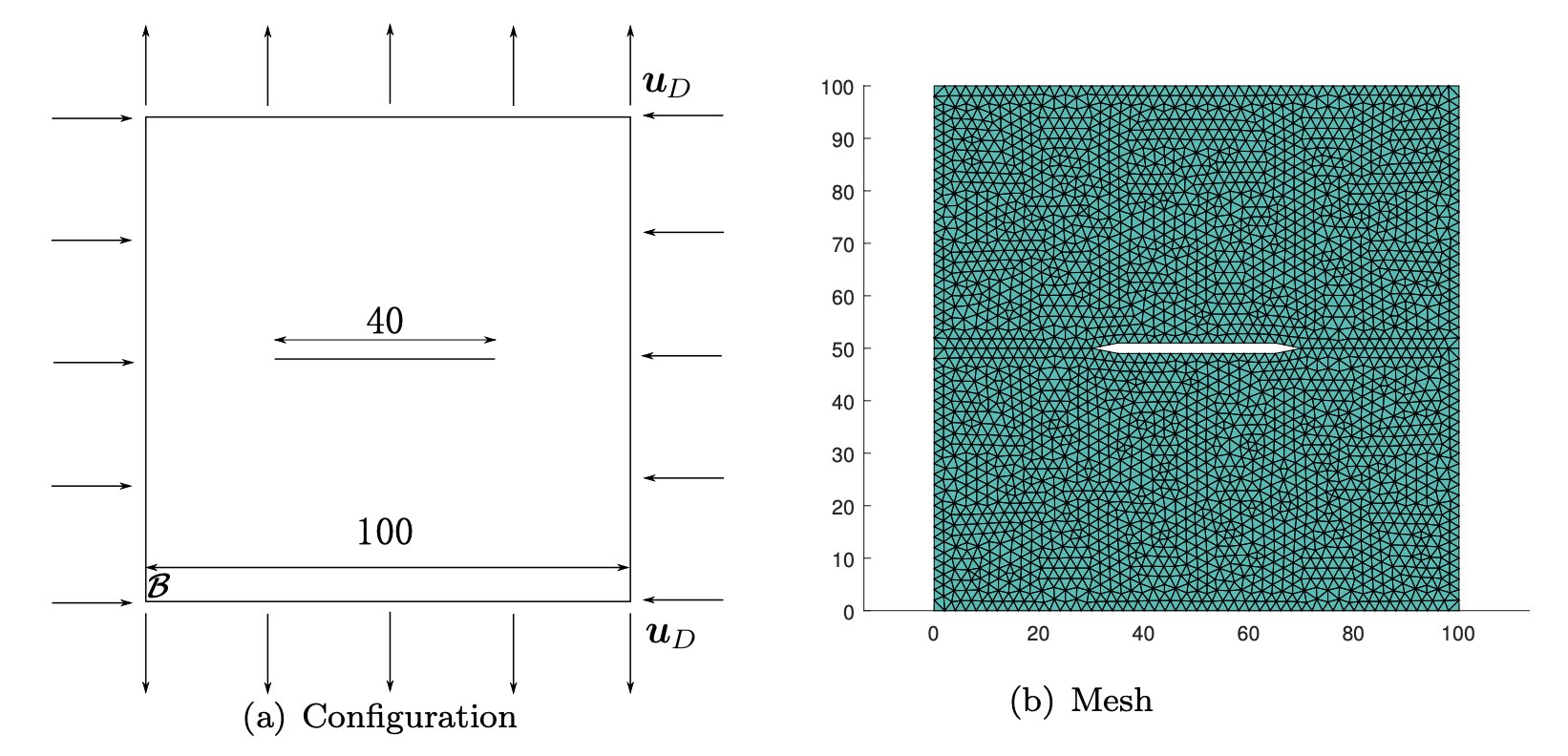}
	\caption{Configuration and mesh of the RVE used for the 2D shear test.}
	\label{decomposition_configuration}
\end{figure}

In order to compare the proposed model with models involving the local crack orientation, i.e., the SS1, SS2, and SK models, 
we perform a shear test for a single RVE with the shear load. The configuration and mesh are shown in Figure \ref{decomposition_configuration}. The material parameters are shown in Table \ref{shear_rotation_material_parameters}.


The RVE domain is $\mathcal{B}=[0,100]^2$. Besides the response of the compared models, we also provide the RVE analysis result from Abaqus for reference. Note that the Abaqus result is based on a calculation with a much finer resolution and which exactly takes into account the self-contact condition due to the crack, see \ref{sec:Abaqus}, while the response from the phase field models are obtained just by a simple function evaluation.


\begin{table}[htbp]
	\centering
	\captionsetup{justification=raggedright,singlelinecheck=false}
	\caption{Material parameters used in the 2D and 3D constitutive response of shear tests}
	\begingroup\setlength{\fboxsep}{0pt}
	\colorbox{lightgray}{%
		\begin{tabular}{ccccc}
			\toprule
			$\lambda$ (GPa) & $\mu$ (GPa) & $r_a$ \\
			\midrule
			$1.1538$ & $0.76923$ & 0.4 \\
			\bottomrule
		\end{tabular}
	}\endgroup
	\label{shear_rotation_material_parameters}
\end{table}

In the shear test, the macroscopic strain is $\overline{\bm{\varepsilon}} = \overline{\varepsilon}_{12}(\bm{e}_1\otimes\bm{e}_2 + \bm{e}_2\otimes\bm{e}_1)$ where $\overline{\varepsilon}_{12}\in[0,1.666\times 10^{-6}]$.

To interpret the results to come, it is helpful to see the simplified expressions of the models under this loading.
The isotropic, V-D, SS1, and SS2 models are simplified to
\[
	\sigma_{12} = g(d)\; 2\mu  \varepsilon_{12}.
\] 

The spectral model is simplified to
\[
	\sigma_{12} = [g(d)+1]\mu\varepsilon_{12}.
\]

The SK model is simplified to
\[
	\sigma_{12} = g(d;b)\; 2\mu \varepsilon_{12}.
\]

The proposed model is simplified to
\[
	\sigma_{12} = g_s(d;\nu)\; 2\mu\varepsilon_{12}.
\]

\begin{figure}[!htbp]
	\centering
	\includegraphics[width=\linewidth]{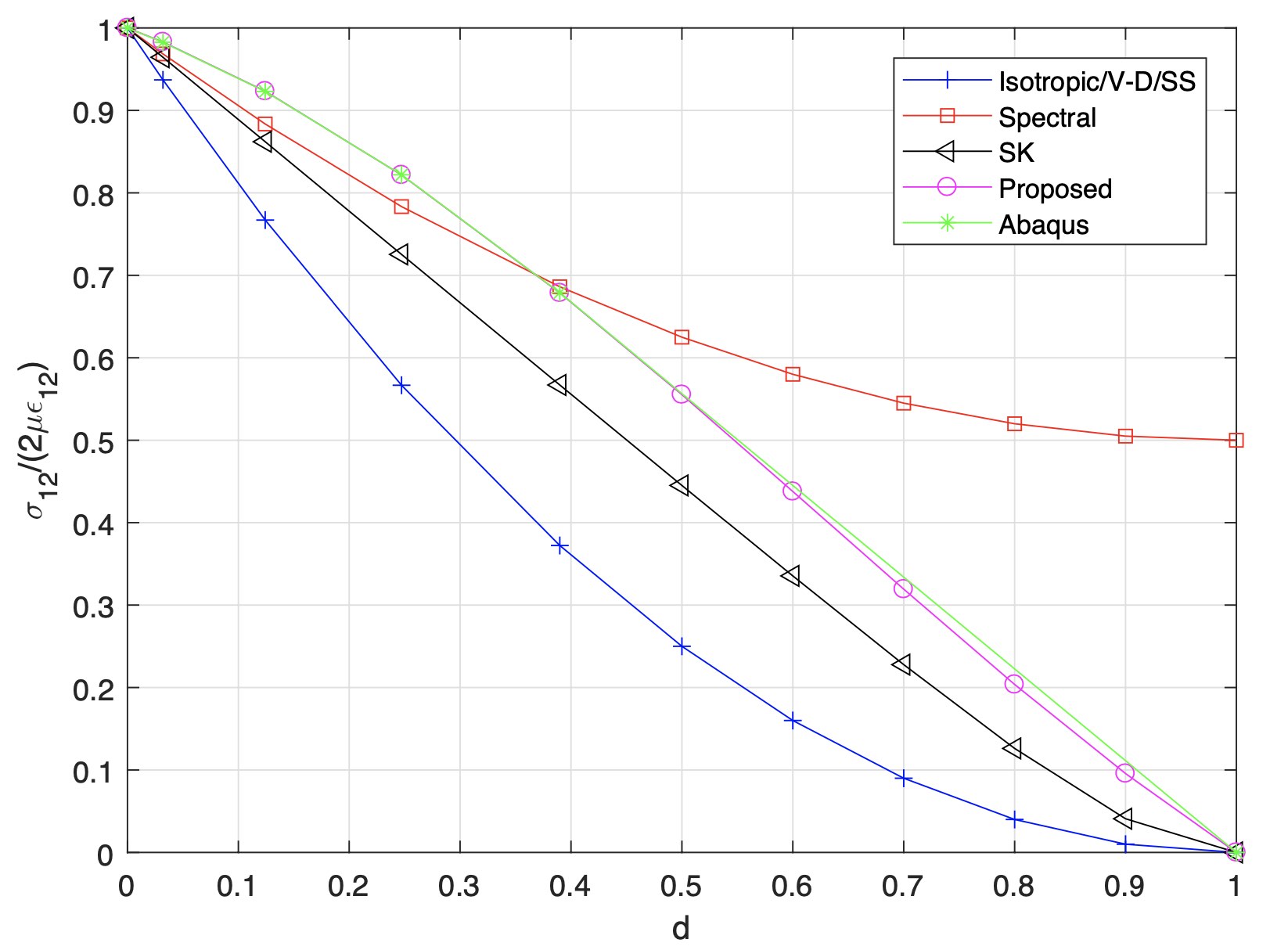}
	\caption{Effective degradation curves $\sigma_{12}/(2\mu\varepsilon_{12})$ vs.~$d$ for the 2D shear test. \texttt{Abaqus} indicates the response of the more costly benchmark solution obtained with Abaqus.}
	\label{sigma12-d}
\end{figure}

In Figure \ref{sigma12-d} we show the curves of $\sigma_{12}/(2\mu\varepsilon_{12})$ vs.~$d$, illustrating the effective shear degradation. Two observations can be made. First, the proposed model is the closest to the benchmark, which is no surprise, as the coefficients in $g_s(d;\nu)$ are obtained by fitting the benchmark solution. Second, it is clear that $\sigma_{12}$ of the spectral model does not degrade to 0 when $d=1$, which explains why the spectral model cannot pass the through crack shear test, see Figure \ref{through_crack_deformed}(b).


\subsection{Constitutive response: 3D combined shear and compression test}
\label{sec:3d_shaer}
\begin{figure}[!htbp]
	\centering
	\includegraphics[width=\linewidth]{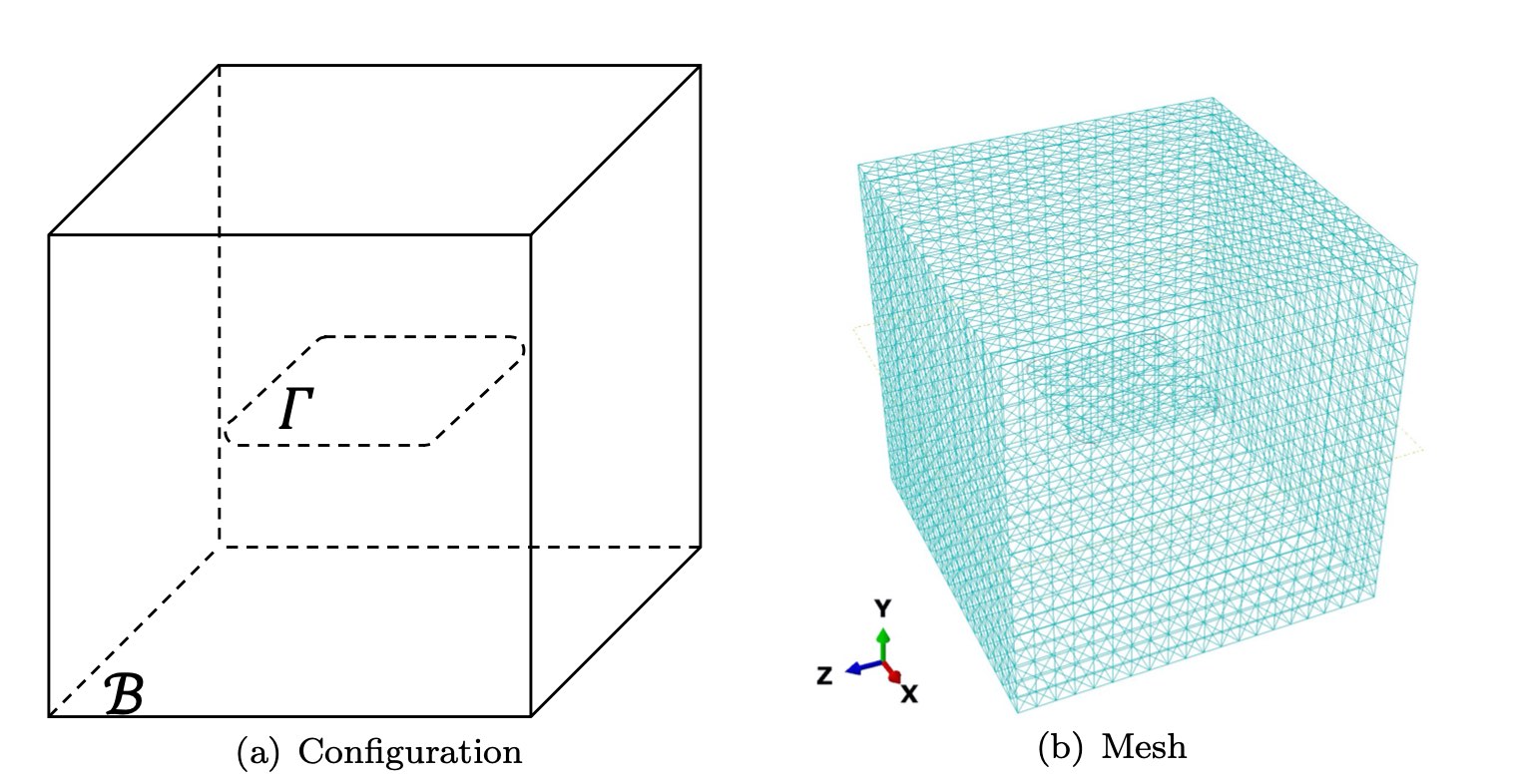}
	\caption{Configuration and mesh of the RVE used for the 3D combined shear and compression test. }
	\label{3d_shear_configuration}
\end{figure}

In the last group of tests, a 3D combined shear and compression load is applied. The configuration and mesh are shown in Figure \ref{3d_shear_configuration}. 
The macroscopic strain is $\overline{\bm{\varepsilon}} = \overline{\varepsilon}_{12}(\bm{e}_1\otimes\bm{e}_2 + \bm{e}_2\otimes\bm{e}_1) + \overline{\varepsilon}_{33}\bm{e}_3\otimes\bm{e}_3$ where $\overline{\varepsilon}_{12}\in[0,1.666\times 10^{-6}]$, $\overline{\varepsilon}_{33}=-1\times 10^{-6}$.

The RVE domain is $\mathcal{B}=[0,100]^3$, and the material parameters are given in Table \ref{shear_rotation_material_parameters}. The same as Section \ref{sec:biaxial}, the Abaqus result takes into account the self-contact condition due to the crack, 
while the response from the phase field models are obtained just by a simple function evaluation.


\begin{figure}[htbp]
	\centering
	\includegraphics[width=\linewidth]{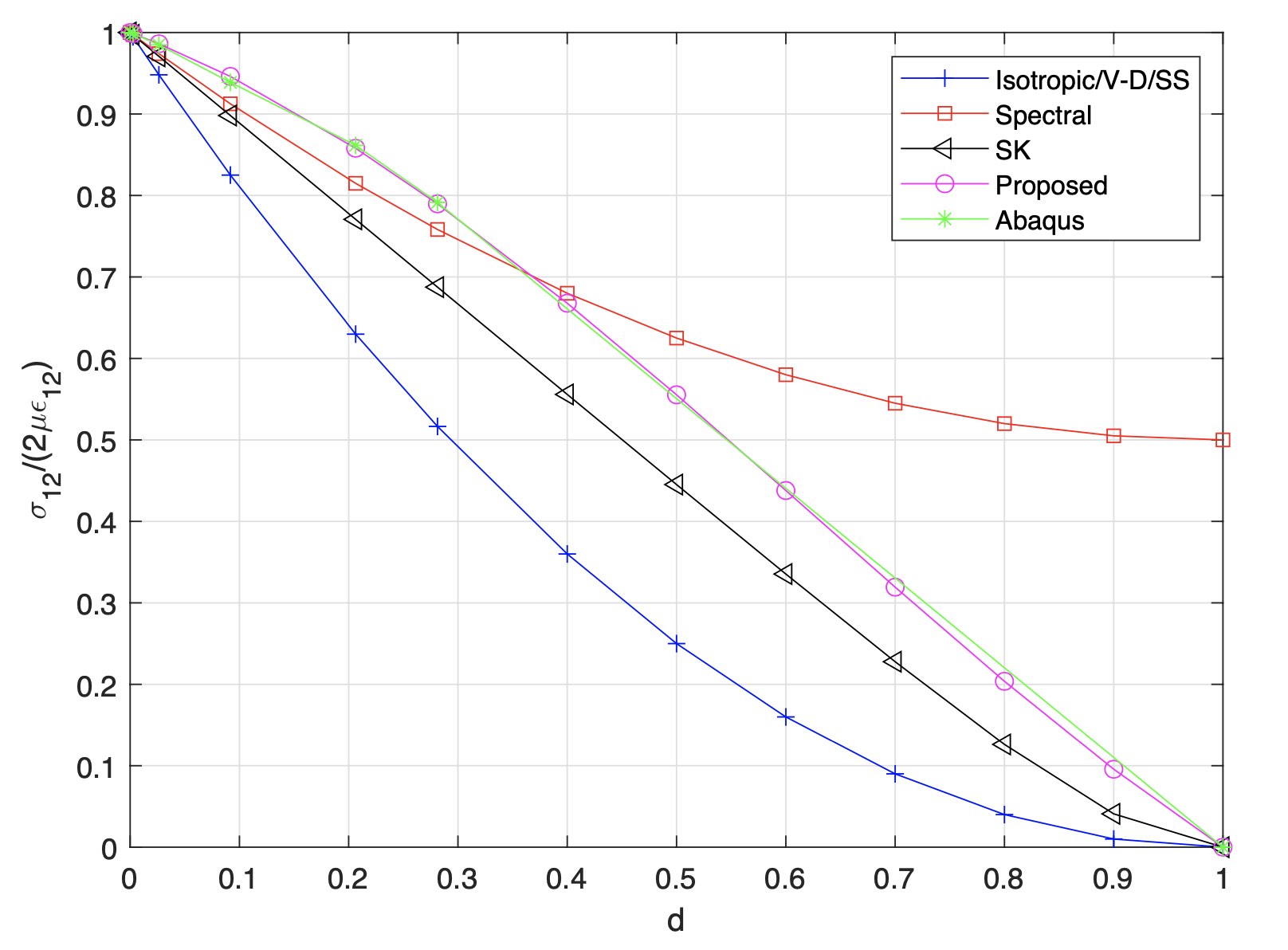}
	\caption{Effective degradation curves $\sigma_{12}/(2\mu\varepsilon_{12})$ vs.~$d$ for the 3D combined shear and compression test. \texttt{Abaqus} indicates the response of the more costly benchmark solution obtained with Abaqus.}
	\label{3d_rve_shear}
\end{figure}

The effective degradation curves $\sigma_{12}/(2\mu\varepsilon_{12})$ vs.~$d$ are shown in Figure \ref{3d_rve_shear}. The proposed model is still the closest to the benchmark Abaqus in this case with mixed three dimensional loads, again, by construction.

\section{Conclusions}
\label{conclusion}

We have proposed a micromechanics-informed phase field model whose macroscopic constitutive relationship is fully determined by the fictitious microstructure. This model possesses the following features:
\begin{enumerate}
	\item The proposed model is based on three shapes of micro-cracks in the RVE. Yet the results from these shapes coincide, once the phase fields are properly calibrated.
	\item The proposed model is the only one that outputs an accurate stiffness response and crack path among the models compared.
	\item The model can be expressed in terms of invariants and pseudo-invariants of the strain tensor, and hence easy to be applied to an arbitrary crack orientation.
	\item The methodology involved in developing the present model can be generalized to more complicated material constitutive relations, such as anisotropy or inelasticty, to be combined with the phase field description of fracture.
\end{enumerate}


\appendix
\section{Details of the numerical experiments on the RVE}\label{sec:Abaqus}

In this appendix, details of the numerical solution of \eqref{RVE-BVP} are provided. This is based on Abaqus/CAE 6.14-2.

\paragraph{Configuration}
We test three types of RVEs: (i) 2D plane strain RVE with a straight crack at the center; (ii) 3D RVE with a penny-shaped crack at the center; (iii) 3D RVE with a square-shaped crack with rounded fillets ($r=0.05L$) at the center. The crack is always perpendicular to the $x_2$-axis.

\paragraph{Mesh}
We use a uniform triangular mesh for the plane strain case for which the element type is \texttt{CPS3} (continuum, plane strain, 3-nodes). And we use a uniform tetrahedral mesh for the 3D cases for which the element type is \texttt{C3D4} (continuum, 3 dimensional, 4-nodes).

\paragraph{Contact condition}
For the frictionless contact conditions, within the contact option \texttt{Tangential Behavior}, we set the \texttt{Friction Coeff} to be 0. This interaction condition is applied to the two sides of the crack $\varGamma$ in RVE. 

\paragraph{Macroscopic stress}
 The macroscopic stress on the RVE is taken as the volumetric average:
\begin{equation*}
	\overline{\bm{\sigma}} = {1\over |\mathcal{B}|}\int_{\mathcal{B}_s } \bm{\sigma} \mathrm{d} V.
\end{equation*}

\paragraph{One-to-one relation of the phase field $d$ and the crack length ratio $r_a$}
We correlate the phase field $d$ with the crack length ratio $r_a$ by calibrating the response of the RVE with uniaxial tensile loads. More precisely, we apply a series of uniaxial tensile loading along the $x_2$-direction, in which case it is assumed that the degradation function $g(d) = (1-d)^2$ is applied to the entire strain energy density of the unbroken material, i.e., \[\varPsi(\bm{\varepsilon},d)=(1-d)^2\varPsi_0(\bm{\varepsilon}).\] 
Substituting $\varepsilon_{11}=\varepsilon_{33}=0$ into \eqref{Eq:sigmas_22} 
yields
\begin{equation}
\frac12 g(d) \left(\lambda+2\mu\right)  \varepsilon_{22}^{2} = \frac{1}{2}  \mathbb{C}^{t}_{2222}\varepsilon_{22}^2, \quad
\forall \varepsilon_y>0.
\label{assumption for d}
\end{equation}

\begin{figure}[htbp]
	\centering
	\includegraphics[width=\linewidth]{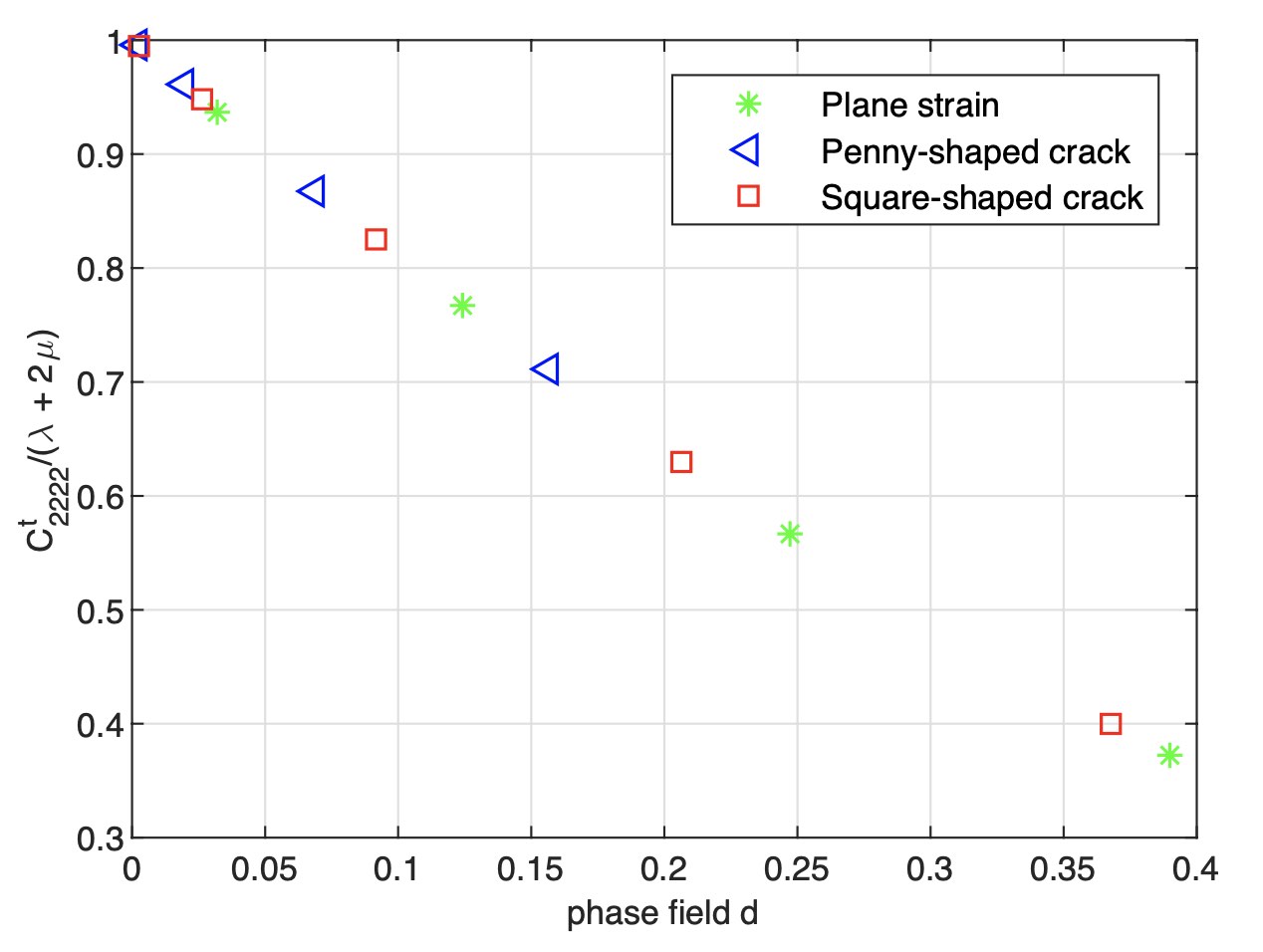}
	\caption{Effective modulus $\mathbb{C}^t_{2222}$ vs.~$d$.}
	\label{c2222_abaqus}
\end{figure}

With \eqref{assumption for d}, the phase field  $d$ 
can then be calculated as
\begin{equation*}
d = 1 - \sqrt{\frac{\mathbb{C}^{t}_{2222}}{\lambda+2\mu}},
\end{equation*}
where $\mathbb{C}^{t}_{2222}$ can be looked up from Figure \ref{c2222_abaqus}.

The phase field $d$ and crack length ratio $r_a$ at $\lambda=1.1538$ GPa and $\mu=0.7692$ GPa (material parameters used by \cite{dascalu2010two}) is given in Table \ref{d-ra}.

\begin{table}[htbp]
	\centering
	\captionsetup{justification=raggedright,singlelinecheck=false}
	\caption{Phase field $d$ at different crack length ratios $r_a$.}
	\begingroup\setlength{\fboxsep}{0pt}
	\colorbox{lightgray}{%
		\begin{tabular}{cccc}
			\toprule
			$r_a$ & $d$ (plane strain) & $d$ (penny-shaped) & $d$ (square-shaped)\\
			\midrule
			0 & 0 & 0 & 0\\
			\midrule
			$0.2$ & $0.0320$ & 0.0021 & 0.0026\\
			\midrule
			$0.4$ & $0.1242$ & 0.0196 & 0.0263\\
			\midrule
			$0.6$ & $0.2472$ & 0.0687 & 0.0917\\
			\midrule
			$0.8$ & $0.3899$ & 0.1566 & 0.2064\\
			\bottomrule
		\end{tabular}
	}\endgroup
	\label{d-ra}
\end{table}

\paragraph{Effective secant moduli}
With both macroscopic strain and stress known, the effective secant moduli can be calculated using the following relationship:
\begin{equation*}
	\overline{\sigma}_{ij} = \mathbb{C}_{ijkl}\overline{\varepsilon}_{kl}.
\end{equation*}

\begin{figure}[htbp]
	\centering
	\includegraphics[width=\linewidth]{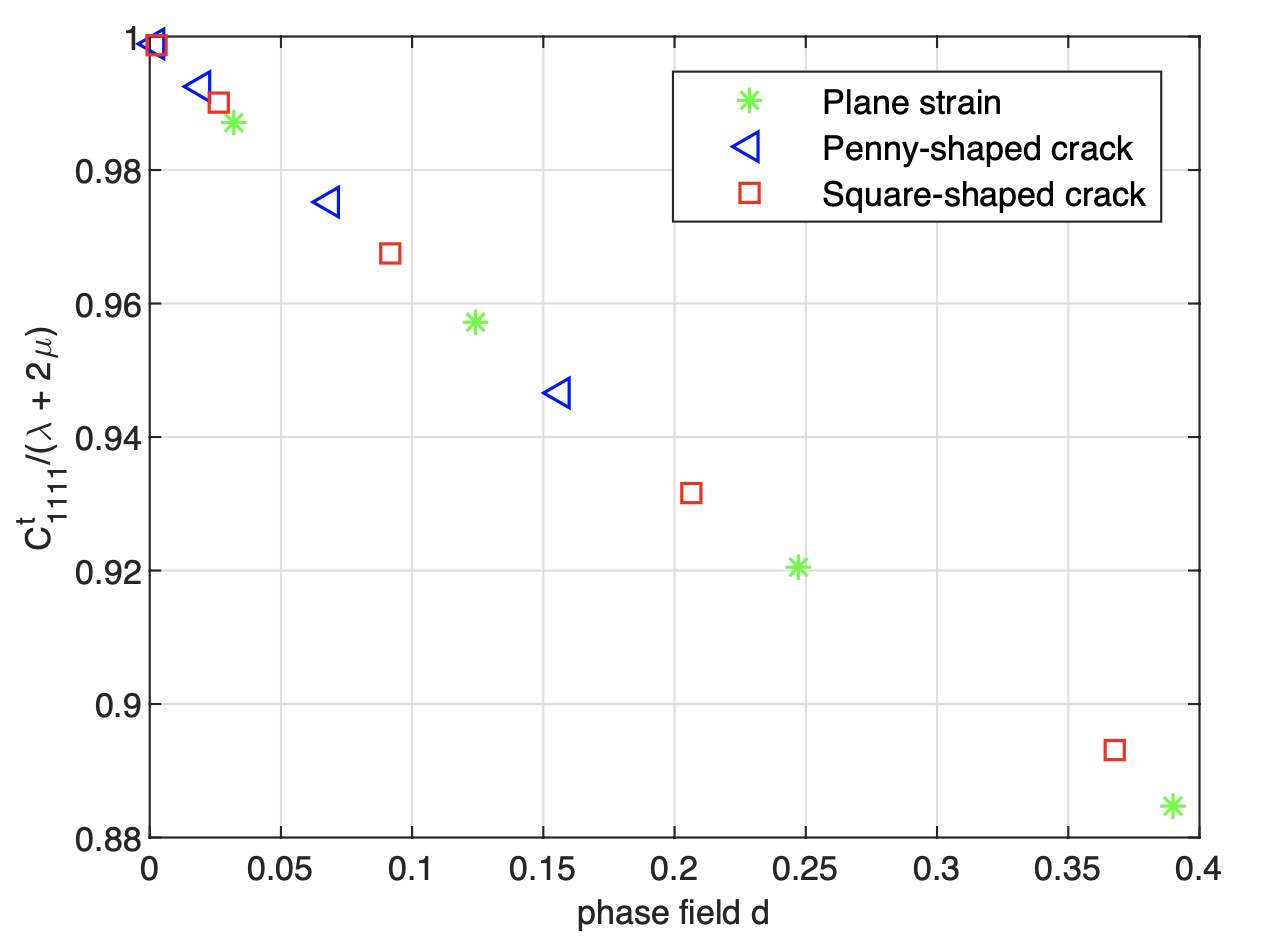}
	\caption{Effective modulus $\mathbb{C}^t_{1111}$ vs.~$d$.}
	\label{c1111_abaqus}
\end{figure}

\begin{figure}[htbp]
	\centering
	\includegraphics[width=\linewidth]{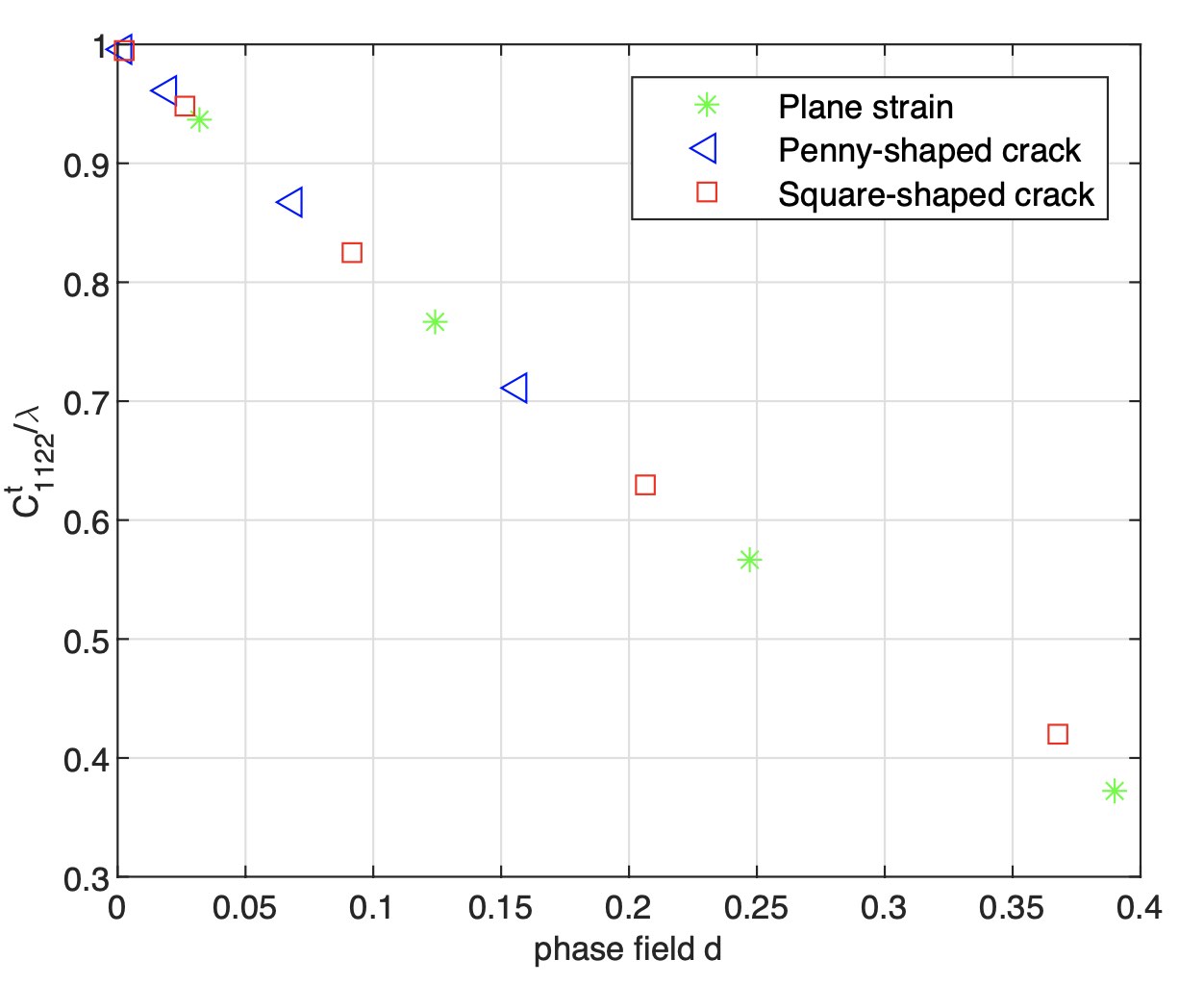}
	\caption{Effective modulus $\mathbb{C}^t_{1122}$ vs.~$d$.}
	\label{c1122_abaqus}
\end{figure}

\begin{figure}[htbp]
	\centering
	\includegraphics[width=\linewidth]{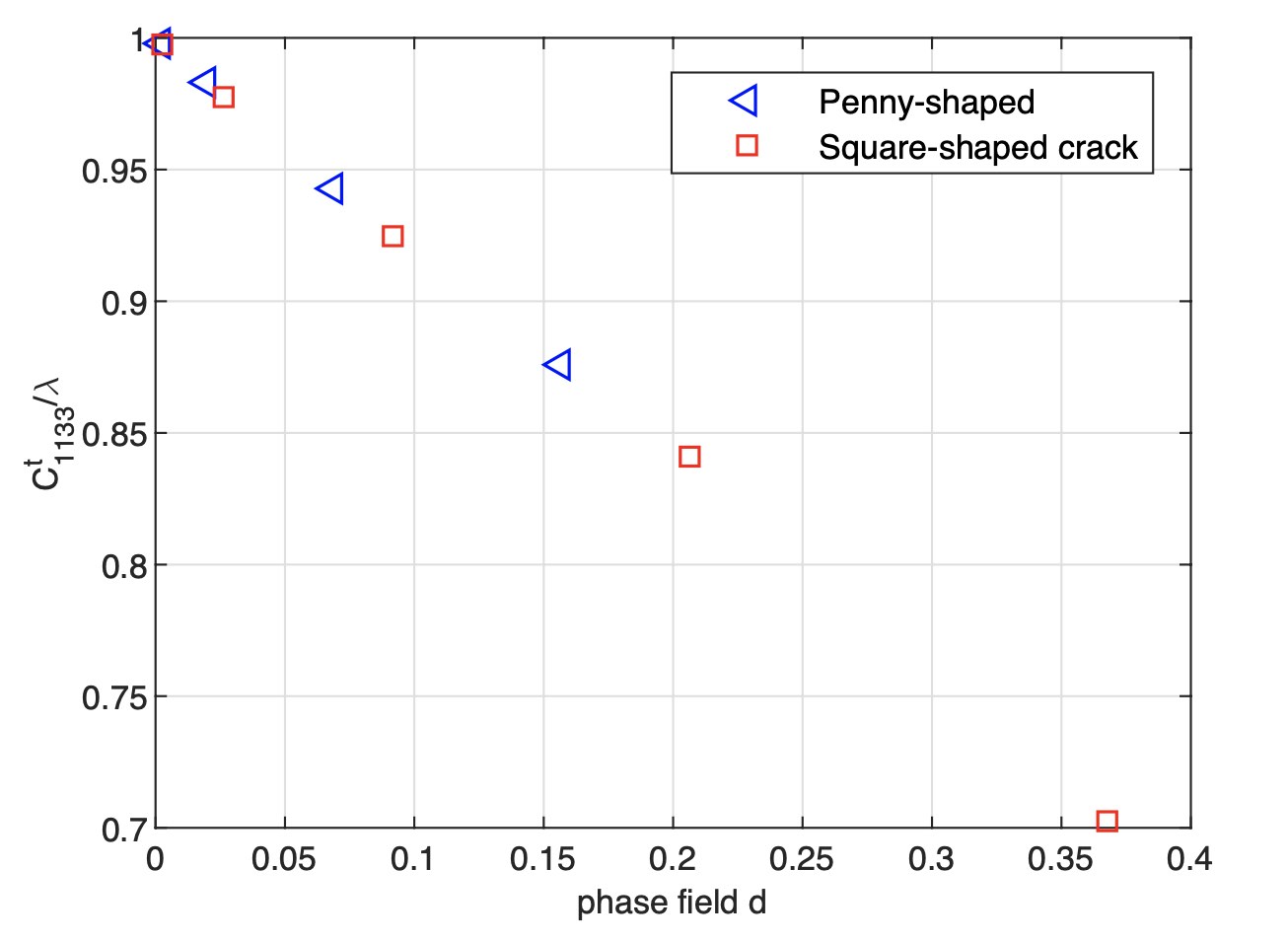}
	\caption{Effective modulus $\mathbb{C}^t_{1133}$ vs.~$d$.}
	\label{c1133_abaqus}
\end{figure}

\begin{figure}[htbp]
	\centering
	\includegraphics[width=\linewidth]{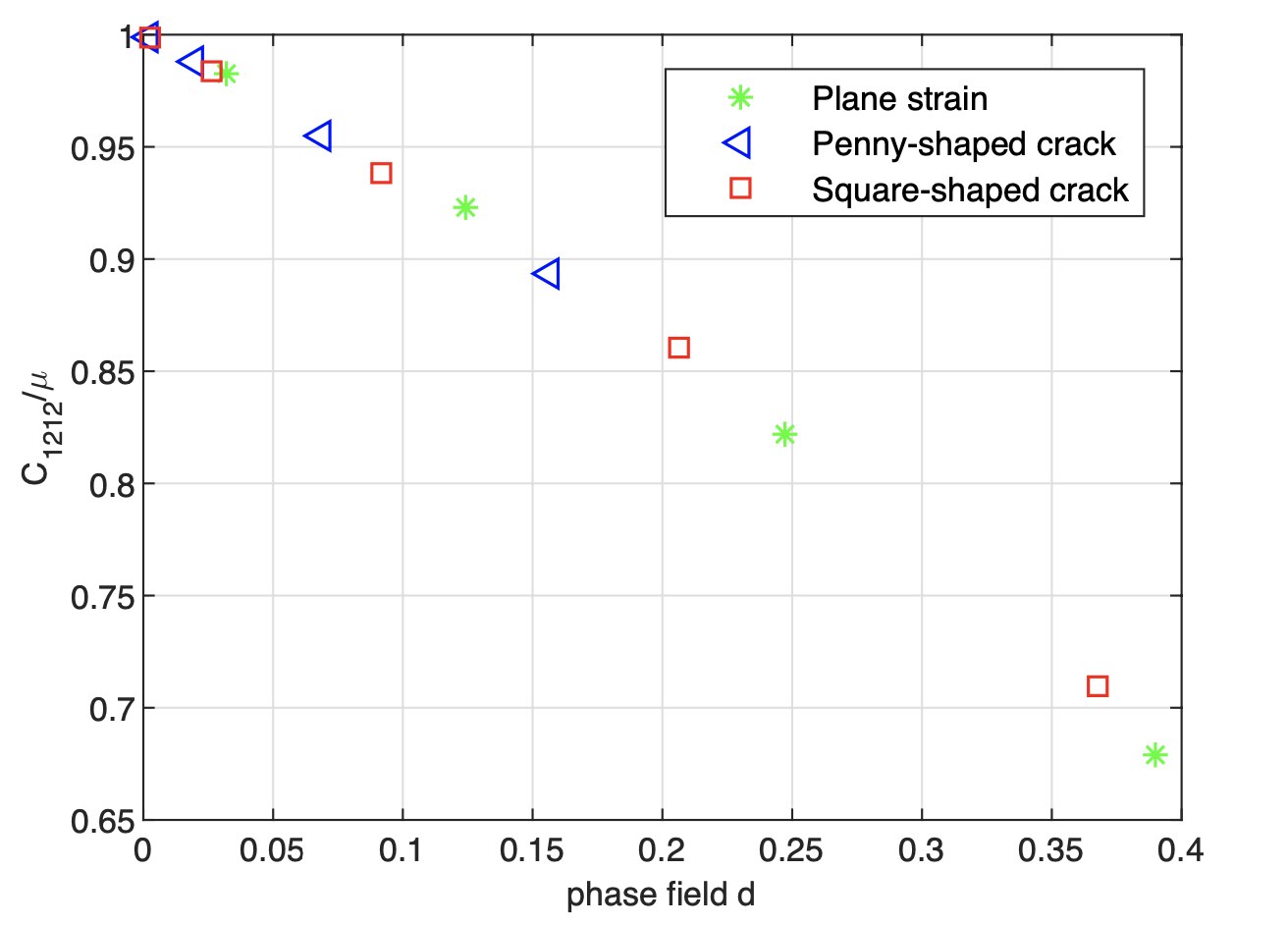}
	\caption{Effective modulus $\mathbb{C}^t_{1212}$ vs.~$d$.}
	\label{c1212_abaqus}
\end{figure}

The numerical results of the effective secant moduli are verified to be independent of $\overline{\bm{\varepsilon}}$ for a given $d$ and given signs of $\beta(\overline{\bm{\varepsilon}})$. The values are shown in Figures \ref{c1111_abaqus}, \ref{c2222_abaqus}, \ref{c1122_abaqus}, \ref{c1133_abaqus}, and \ref{c1212_abaqus}.

\section{Deriving the model in terms of invariants}
\label{sec:invariants}
The key step in the derivation is to express the strain components in terms of the (pseudo-)invariants.
We first choose a very special coordinate system such that the $x_2$-axis is normal to the crack $\varGamma$, and that $\varepsilon_{13}=0$. In this special case, it is easy to deduce that
\[
\varepsilon_{22}=I_4,\quad \varepsilon_{12}^2 + \varepsilon_{23}^2 = I_5 - I_4^2,
\quad
\varepsilon_{11}+\varepsilon_{33} = I_1 - I_4,
 \quad
\varepsilon_{11} \varepsilon_{33} = I_2 + I_5 - I_1 I_4.
\]
Applying these expressions in \eqref{special-orientation} yields \eqref{with-invariants}.

As a sanity check, we now still take the $x_2$-axis normal to $\varGamma$, and $\bm{\varepsilon}$ to have all zero components except $\varepsilon_{13}=\varepsilon_{31}$. Then $I_1=I_4=I_5=0$, and $I_2=-\varepsilon_{13}^2$. It can be shown that $\varPsi_t=\varPsi_c=2\mu\varepsilon_{13}^2$ while $\varPsi_s=0$, giving rise to the expected expression, $\varPsi=2\mu\varepsilon_{13}^2$.


\section*{Acknowledgments}
This work is supported by the Shanghai Municipal Science and Technology Commission, Grant No.~19ZR1424200, and by the National Natural Science Foundation of China, Grant No.~11972227.







\bibliographystyle{elsarticle-num}
\bibliography{Arxiv_MI}

\end{document}